\documentclass[12pt]{article}
\usepackage{a4,amsmath,theorem,times,euler,euscript,epsfig,multirow}
\renewcommand{\baselinestretch}{1.1}
\theoremstyle{plain}\theorembodyfont{\sl}%
                    \newtheorem{principle}{Principle}
\newcommand{\Principle}[1]{Principle~\ref{#1}}
\theoremstyle{break}\theorembodyfont{\rmfamily}%
                    \newtheorem{algorithm}{Algorithm}
\newcommand{\Algorithm}[1]{Algorithm~\ref{#1}}
\newcounter{contstep}
\newcommand{\initstep}{\setcounter{contstep}{0}}
\newcommand{\addstep}{\addtocounter{contstep}{+1}}
\newcommand{\additem}{\addstep\item}
\newcommand{\addsteps}{\setcounter{enumi}{\thecontstep}}
\newcommand{\Appendix}[1]{Appendix~\ref{#1}} 
\newcommand{\Section}[1]{Section~\ref{#1}}
\newcommand{\Figure}[1]{Figure~\ref{#1}}

\newcommand{\df}{:=}

\newcommand{\vhi}{\varphi}
\newcommand{\ve}{\varepsilon}
\newcommand{\srac}[2]{{\textstyle\frac{#1}{#2}}}
\newcommand{\grac}[2]{#1/#2}
\newcommand{\apriori}{{\it a priori\/}}
\newcommand{\apd}{a.p.d.}
\newcommand{\upd}{u.p.d.}
\newcommand{\upds}{u.p.d.s}
\newcommand{\Dap}{{\boldsymbol\omega}}
\newcommand{\dap}{\omega}
\newcommand{\Pdns}{P}
\newcommand{\meas}[1]{\langle\,#1\,\rangle}
\newcommand{\Pmeas}[1]{\langle\,#1\,\rangle_{\Pdns}^{}}
\newcommand{\Omeas}[1]{\langle\,#1\,\rangle_{\Dap}^{}}
\newcommand{\nda}{N}
\newcommand{\dspace}{\Omega}
\newcommand{\we}{x}
\newcommand{\Gdns}[1]{G_{#1}}
\newcommand{\gdns}{g}
\newcommand{\nwe}{n}
\newcommand{\Ent}{H}
\newcommand{\lag}{\lambda}
\newcommand{\Ord}{\EuScript{O}}
\newcommand{\lar}{\leftarrow}

\newcommand{\vol}{v}
\newcommand{\ind}{\vartheta}
\newcommand{\dimO}{s}
\newcommand{\parni}{{\tt PARNI}}
\newcommand{\dirac}{\delta}
\newcommand{\scut}{s_{\mathrm{c}}}
\newcommand{\Real}{\mathbf{R}}
\newcommand{\rambo}{{\tt RAMBO}}
\newcommand{\hicom}{{\tt HICOM}}
\newcommand{\haag}{{\tt HAAG}}
\newcommand{\sarge}{{\tt SARGE}}
\newcommand{\Ant}{A}
\newcommand{\Asym}{A_{\mathrm{sym}}}
\newcommand{\Nge}{\nda}
\newcommand{\Nac}{\nda_{\mathrm{acc}}}
\newcommand{\Ttot}{T_{\mathrm{tot}}}
\newcommand{\Tint}{T_{\mathrm{int}}}
\newcommand{\stdev}{\sigma}
\newcommand{\nl}{\notag\\}
\begin{document}
\begin{center}
%
{\bf\huge Adaptive channels for data analysis}\\\vspace{0.25\baselineskip}
{\bf\huge and importance sampling}

\vspace{2\baselineskip}
{\Large Andr\'e van Hameren}

\vspace{0.25\baselineskip}
{\large\it Institute of Nuclear Physics, NCSR Demokritos, Athens, Greece}
{\large\tt andrevh@inp.demokritos.gr}

\vspace{0.25\baselineskip}
{\large\today}

\renewcommand{\baselinestretch}{1}
\vspace{2\baselineskip}
{\bf Abstract}\\\vspace{0.5\baselineskip}
\parbox{0.8\linewidth}{\small\hspace{15pt}%
The adaptive multi-channel method is applied to derive probability
distributions from data samples. Moreover, an explicit algorithm is
introduced, for which both the channel weights and the channels themselves are
adaptive, and which can be used both for data analysis and for importance
sampling in Monte Carlo integration. Finally, it is pointed out how the
usefulness for data analysis can be used to optimize the integration procedure.
}
\end{center}
\vspace{0.5\baselineskip}
{\sl PACS\/}: 02.60.Pn; 02.70.Lq.\\
{\sl keywords\/}: adaptive Monte Carlo integration; data analysis.
\vspace{0.5\baselineskip}

\section{Introduction}
Since its introduction in \cite{KleissPittau}, adaptive multi-channeling has
been extensively used as a help for importance sampling in numerical
integration. The probability density, following which the integration points
are generated, is written as a sum of densities with positive weights, and these
weights are optimized during the integration process, in order to minimize the
variance. The densities in the sum are called {\em channels}. In this paper, 
two variations of this method are introduced.

Firstly, we observe the similarity between the optimization of the density
following which the integration points are generated, and the creation of a
histogram in data-analysis. A histogram is a weighted sum of densities, given
by the normalized non-overlapping indicator functions of subsets of the space
in which the data-points take their values. These indicator functions are the
`bins'. The optimal weights are estimated by the number of data points in the
bins, and these are exactly the maximum likelihood estimators. In this paper,
we will see that a histogram is a special case of the result of the application
of a multi-channel method to derive the probability distribution. This will
lead to `generalized histograms' or {\em unitary probability decompositions\/}
(\upds), by the use of general probability densities instead of the indicator
functions. In \Section{sec1}, it will be pointed out how the weights can be
optimized.

Secondly, we note that the adaptation in the application of the multi-channel
method to importance sampling has mainly been applied to the {\em weights\/} in
the sum of weighted densities.  The channels themselves are fixed, which
presupposes some knowledge about the integrand based upon which the particular
channels have been chosen.  If one does not want to rely on such knowledge to
much, one can try to adapt also the channels.  The simplest way to do this is
by discarding channels with low weight during the integration process, or
choosing new sets of channels from a larger given pool of
channels~\cite{vanHamerenPapadopoulos}. A more advanced way of channel
adaptation would be in the spirit of the {\tt VEGAS}-algorithm~\cite{Lepage} or
the {\tt FOAM}-algorithm~\cite{Jadach}, by creating new channels based on
statistical analyses of the existing channels.  The disadvantage of the first
algorithm is that it is of limited efficiency in more than one dimension if the
integrand has non-factorizable peak structures. The disadvantage of the second
algorithm is that its complexity increases factorially with the dimension of
the integration space. The problem with {\tt VEGAS} can be solved if some
information about the integrand is available, through the use of various
channels corresponding with different coordinate systems~\cite{Ohl}.

In \Section{sec3}, 
the algorithm \parni\
will be introduced. It uses adaptive multi-channeling,
with channels that are fully adaptive themselves. 
It has no \apriori\ problems with non-factorizable peak structures in the 
integrand, and its complexity grows linearly with the dimension of the
integration space.  Except for automatic importance sampling in Monte Carlo
integration, \parni\ can also be used for 
the creation of a \upd. In \Section{sec4} we will see how this property 
can be used to optimize the integration process, and this strategy will be 
applied to a problem in phase space integration.

\section{Multi-channeling for data analysis\label{sec1}}
The derivation of a probability distribution from a sample of data is a common
problem in scientific research.  The idea is that such a distribution exists
\apriori, and that the data are drawn at random. 
The Bayesian interpretation then tells us that the \apriori\
distribution (\apd) gives the probability for this particular sample of data to
be drawn.  The frequentist interpretation tells us that estimators, calculated
with the
sample, will converge to the same values as the ones from the \apd\
if the sample becomes very large.  This interpretation can be translated into
the statement that the distribution of the sample converges to the \apd\
if the sample becomes large.

An obvious way to derive this distribution is by using such estimators, and the
assumption that the distribution belongs to a class that can be completely
described by the parameters that are estimated.
In most cases, the original problem is even stated such, that the distribution
is of a class that can be completely described by a set of parameters, and that
one only wants to determine the correct values of these parameters.
Part of the problem then becomes the determination of the right 
estimators, for example the maximum likelihood.

One particular kind of estimators are the number of data points in subsets
of the space in which the data points take their values.
Such a number, divided by the total number of data points, estimates the
integral of the \apd\
over the subset.  The estimates for a collection of non-overlapping subsets, or
{\em bins\/}, that cover the whole space constitute a {\em histogram\/}, and in
certain limits in which the number of bins and the number of data go to
infinity, one can speak about convergence of the histogram to the probability
density of the \apd.

A histogram is a weighted sum of densities, given by the normalized indicator
functions of the subsets.
The estimators described above are exactly the maximum likelihood estimators 
of these weights.
In this section, we will see that a histogram is a special case of the result
of the application of the multi-channel method to derive the probability
distribution, and we will see how to generalize it to a {\em unitary
probability decomposition\/} (\upd), by the use of general probability
densities instead of the indicator functions.  

\subsection{Maximum likelihood and entropy}
If data points are assumed to be distributed following a probability density
which is given up to the values of certain parameters, the maximum likelihood
gives estimators for these parameters. For our
application, it will be more convenient to formulate the maximum likelihood
method in terms of the maximization of an entropy.
The space in which data points
$\dap$
take their values is denoted
$\dspace$,
and this space may be multi-dimensional.
The entropy of a probability density
$\Pdns$
on
$\dspace$
is given by 
\begin{displaymath}
   \Ent(\Pdns) = \int_{\dspace}\Pdns(\dap)\log\Pdns(\dap)\,d\dap
\;\;.
\end{displaymath} 
It can be used to determine the probability distributions of a system for which
some information is available. 
This information could be the values of characteristics like mean and variance,
but could also be knowledge that 
$\Pdns=\Gdns{\we}$,
where
$\Gdns{\we}$
is given up to the values of the parameters
$\we=(\we_1,\we_2,\ldots,\we_{\nwe})$.
The shape of
$\Pdns$,
or the values of
$\we$,
should be such that the entropy assumes its maximum value
\footnote{Or such that $-\Ent(\Pdns)$ assumes its minimal value.}.

But we are interested in a different situation, with a
sample or a stream of data as the only information available.
So the \apriori\
probability density
$\Pdns$
is given somehow, and we do not know it, but want to approximate is with the
help of the data, using a parametrized density
$\Gdns{\we}$,
specially chosen for this task.
$\Pdns$
defines a probability measure on
$\dspace$,
and for a measurable function
$f$
we write
\begin{displaymath}
   \Pmeas{f} \df \int_{\dspace}f(\dap)\,\Pdns(\dap)d\dap
\;\;.
\end{displaymath}
In search for the optimal values of 
$\we$,
we introduce the entropy of
$\Gdns{\we}$
relative to
$\Pdns$:
\begin{equation}
   \Ent(\Pdns;\Gdns{\we})
   \df \Pmeas{\log\Gdns{\we}}
\;\;.
\label{defent2}\end{equation}
It is the original entropy with
$\log\!\Pdns$
replaced by
$\log\!\Gdns{\we}$,
and naturally, one would expect that if
$\we$
is optimal, then the two entropies are close together.
Encouraged by this expectation, we state that
\begin{principle}
  $\we$ 
  is optimal if\/ 
  $\Ent(\Pdns;\Gdns{\we})$ 
  assumes its maximum value.
\label{princ1}\end{principle}
Extrema of
$\Ent(\Pdns;\Gdns{\we})$
are given by solutions of the equations
\begin{displaymath}
   0 = \frac{\partial}{\partial\we_i}\,\Ent(\Pdns;\Gdns{\we}) 
     = \frac{\partial}{\partial\we_i}\Pmeas{\log\Gdns{\we}}
\;\;,\quad 
   i=1,\ldots,\nwe
\;\;.
\end{displaymath}
The connection with the maximum likelihood can be established by realizing
that, in real life, one does not know
$\Pdns$,
so that integrals over
$\dspace$
have to be estimated with the help of the available data, which are
distributed following
$\Pdns$.
In fact, if
$\Dap=(\dap_1,\dap_2,\ldots,\dap_{\nda})$
is such a sample of data points, and if
$\Pmeas{f^2}$
exists, then
\begin{displaymath}
   \Omeas{f}
   \df \frac{1}{\nda}\sum_{k=1}^{\nda}f(\dap_k)
   \overset{\nda\rightarrow\infty}{\longrightarrow} \Pmeas{f}
\;\;,
\end{displaymath} 
where convergence takes place at least in probability, like in Monte Carlo 
integration. 
Using this estimator for the integral, the equations become
\begin{displaymath}
   0 = \nda\frac{\partial}{\partial\we_i}\Omeas{\log\Gdns{\we}}
     = \frac{\partial}{\partial\we_i}\sum_{k=1}^{\nda}\log\Gdns{\we}(\dap_k)
     = \frac{\partial}{\partial\we_i}\log\prod_{k=1}^{\nda}\Gdns{\we}(\dap_k)
\;\;,
\end{displaymath}
so that solutions give extrema of the likelihood function.
We prefer to stick to the entropy formulation from now on, because it seems
more appropriate in the case of a continuous data stream, a situation that
resembles the one of Monte Carlo integration, where a continuous stream of data
points is generated in order to integrate a function.

\subsection{Entropy and multi-channeling}
From now on,
$\Gdns{\we}$
will always be linear in the parameters
$\we=(\we_1,\we_2,\ldots,\we_{\nwe})$,
and be defined with the help of
$\nwe$
probability densities
$\gdns_i$,
or {\em channels\/},
by
\begin{equation}
   \Gdns{\we}(\dap) = \sum_{i=1}^{\nwe}\we_i\gdns_i(\dap)
\;\;.
\label{defGdns}\end{equation}
The parameters, or {\em weights\/} will always be positive, and normalized such
that
$\sum_{i=1}^{\nwe}\we_i=1$.
If we look for extrema of the entropy (\ref{defent2}), we have to take care
that the weights stay normalized, and instead of including this normalization
in 
$\Gdns{\we}$
explicitly, we prefer to extend the entropy with the help of a Lagrange
multiplier. 
So we want to find the maximum of 
\begin{displaymath}
   \Ent(\Pdns;\Gdns{\we},\lag)
   \df \Pmeas{\log\Gdns{\we}} - \lag\Big(1-\sum_{i=1}^{\nwe}\we_i\Big)
\end{displaymath}
with respect to
$\we$
{\em and\/}
$\lag$.
Extrema are solutions to the equations
\begin{displaymath}
   \Pmeas{\grac{\gdns_i}{\Gdns{\we}}} = \lag
   \;\;, \quad i=1,\ldots,\nwe
\qquad\textrm{and}\qquad
   \sum_{i=1}^{\nwe}\we_i = 1
\;\;.
\end{displaymath}
The value of the Lagrange multiplier
$\lag$
can be found by multiplying the equation
$\Pmeas{\grac{\gdns_i}{\Gdns{\we}}} = \lag$
with
$\we_i$
and taking the sum over
$i$.
Remembering (\ref{defGdns}), we then find that
$\lag=1$.

The question is now whether this solution corresponds to a maximum, 
and performing an analysis like in~\cite{KleissPittau}, we can establish that 
it leads to at least a local maximum: denoting the solution by 
$\Bar{\we}$
and taking a small variation
$\ve=(\ve_1,\ldots,\ve_{\nwe})$
with
$\sum_{i=1}^{\nwe}\ve_i=0$,
around the solution we find
\begin{displaymath}
   \Ent(\Pdns;\Gdns{\Bar{\we}+\ve},1)
   =  \Ent(\Pdns;\Gdns{\Bar{\we}},1)
     -\frac{1}{2}\Pmeas{\grac{\Gdns{\ve}^2}{\Gdns{\Bar{\we}}^2}}
     +\Ord(\ve^3)
\;\;,
\end{displaymath}
and we see that small variations lead to a decrease of the entropy.
We can re-formulate \Principle{princ1} now, and state that
if 
$\Gdns{\we}$
is a weighted sum of channels
$\gdns_i$,
then 
\begin{principle}
  $\we$ 
  is optimal if the\/ 
  ${\displaystyle \Pmeas{\grac{\gdns_i}{\Gdns{\we}}}}$
  are equal (to\/ $1$) for all\/
  $i=1,\ldots,\nwe$.
\label{princ2}\end{principle} 

\subsection{Numerical path to the solution}
In general it is difficult to find an analytic solution to the problem posed
by \Principle{princ2}.
We can, however, try to find the solution, or at least an approximation, by
numerical methods.
We follow the same path as in~\cite{KleissPittau} 
(\Appendix{VarOpt}), by considering the case in 
which the channels
$\gdns_i$
are normalized indicator functions of non-overlapping subsets of
$\dspace$
with volume
$\vol_i$,
so that
$\Gdns{\we}$
is a histogram.
Let 
$\ind_i=\vol_i\gdns_i$
denote the indicator functions.
Then 
\begin{displaymath}
   \Pmeas{\grac{\gdns_i}{\Gdns{\we}}}
   = \frac{1}{\we_i}\Pmeas{\ind_i}
\quad\textrm{so that the solution is given by}\quad
   \we_i = \Pmeas{\ind_i}
\;\;.
\end{displaymath} 
So the optimal weight of channel
$\gdns_i$
is given by to the integral of the probability density over the subset
corresponding to the indicator function.
The weight for the indicator function is then given by
$\we_i/\vol_i$:
the height of the bin, like usually for a histogram.
We observe now that, starting from any
$\we$, 
the successive operations
\begin{algorithm}[unitary probability decomposition]
\begin{enumerate}
\item $y_i 
       \lar \we_i{\displaystyle\Pmeas{\grac{\gdns_i}{\Gdns{\we}}}}$
      \hspace{10pt}for all $i=1,\ldots,\nwe$
\item $\we_i \lar {\displaystyle\frac{y_i}{\sum_{j=1}^{\nwe}y_j}}$
      \hspace{10pt}for all $i=1,\ldots,\nwe$
\end{enumerate}
\label{alg1}\end{algorithm}
lead directly to the optimal values for
$\we$,
and this gives us faith to seek for the solution in the general case by 
recursive application of these operations.

In real life again, we cannot calculate
$\Pmeas{\grac{\gdns_i}{\Gdns{\we}}}$,
and have to estimate it with an available sample 
$\Dap$
by
$\Omeas{\grac{\gdns_i}{\Gdns{\we}}}$.
Notice that, in the case that
$\Gdns{\we}$
is a histogram,
$\we_i\Omeas{\grac{\gdns_i}{\Gdns{\we}}}=\Omeas{\ind_i}$
is the number of data points in bin $i$.

\subsection{An example of unitary probability decompositions}
\begin{figure}
\begin{center}
\epsfig{figure=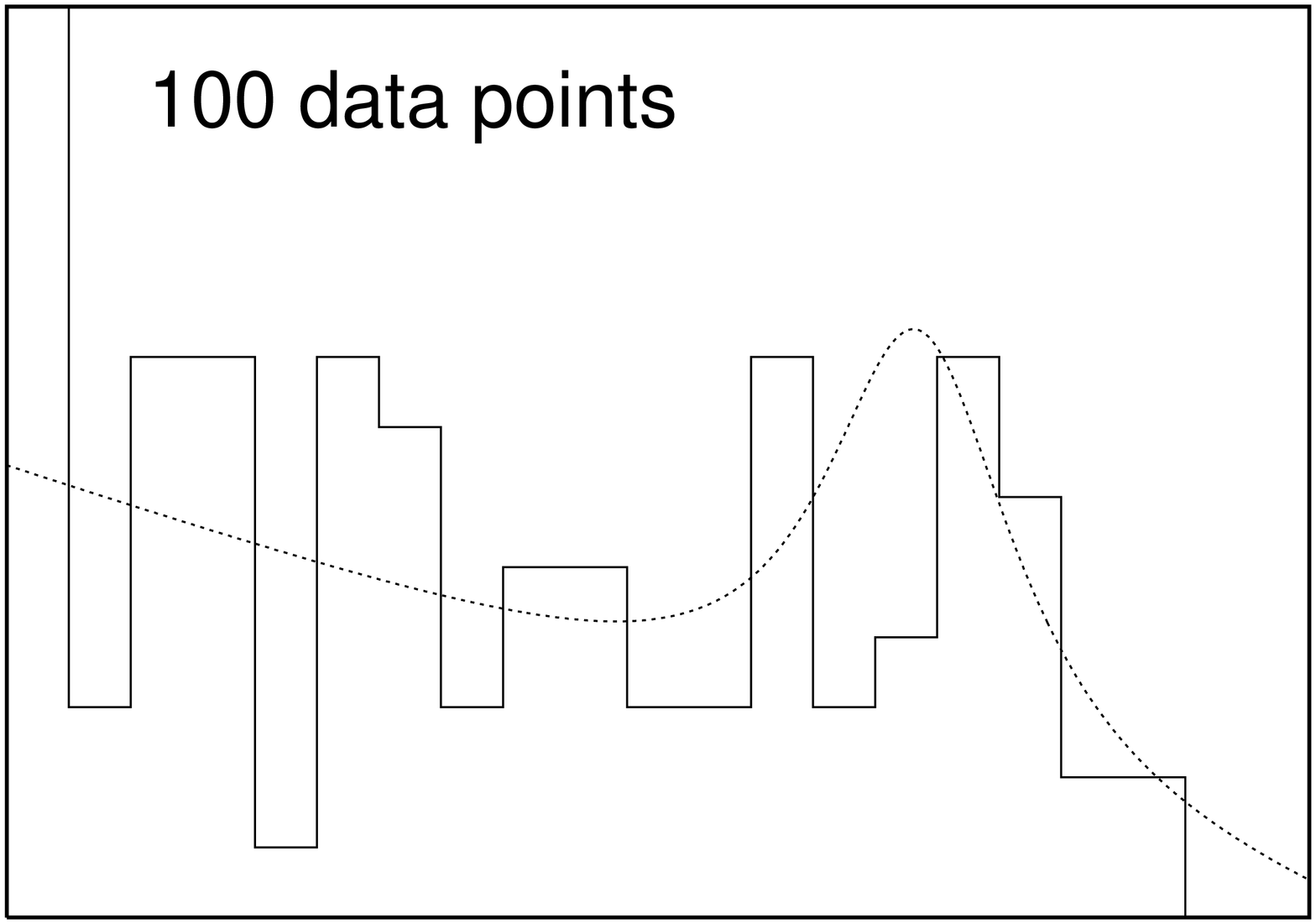,width=0.23\linewidth,angle=270}%
\epsfig{figure=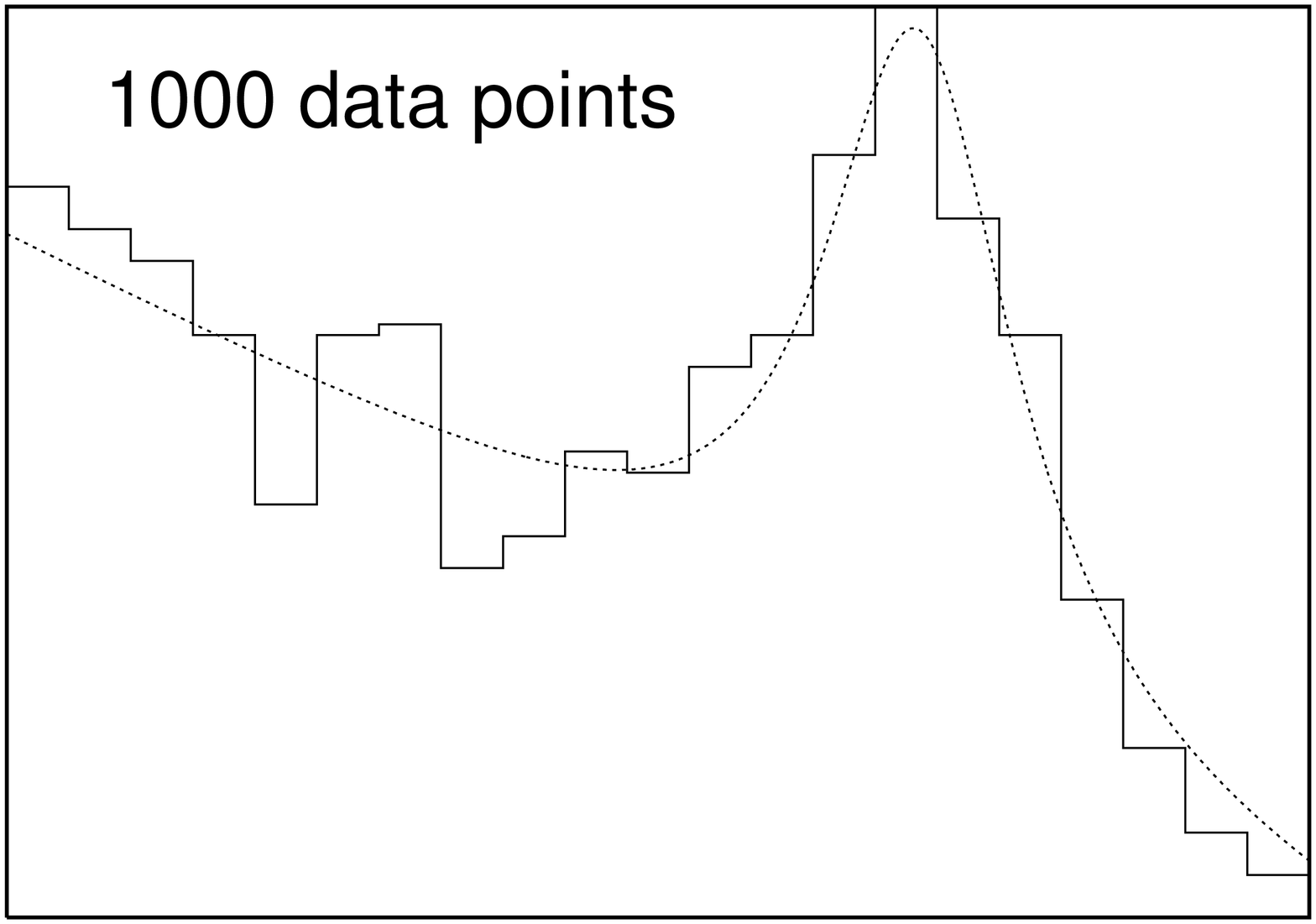,width=0.23\linewidth,angle=270}%
\epsfig{figure=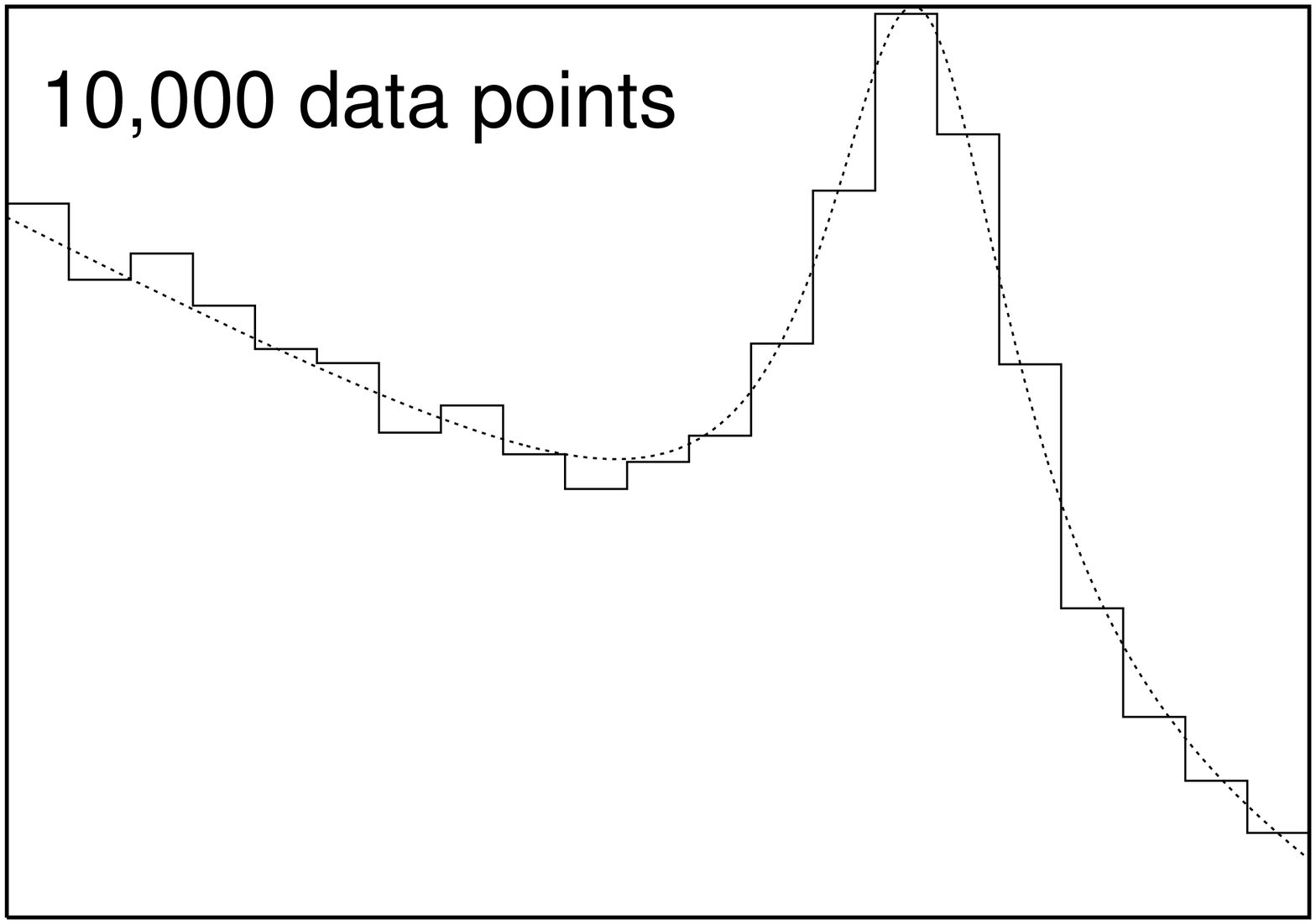,width=0.23\linewidth,angle=270}
\epsfig{figure=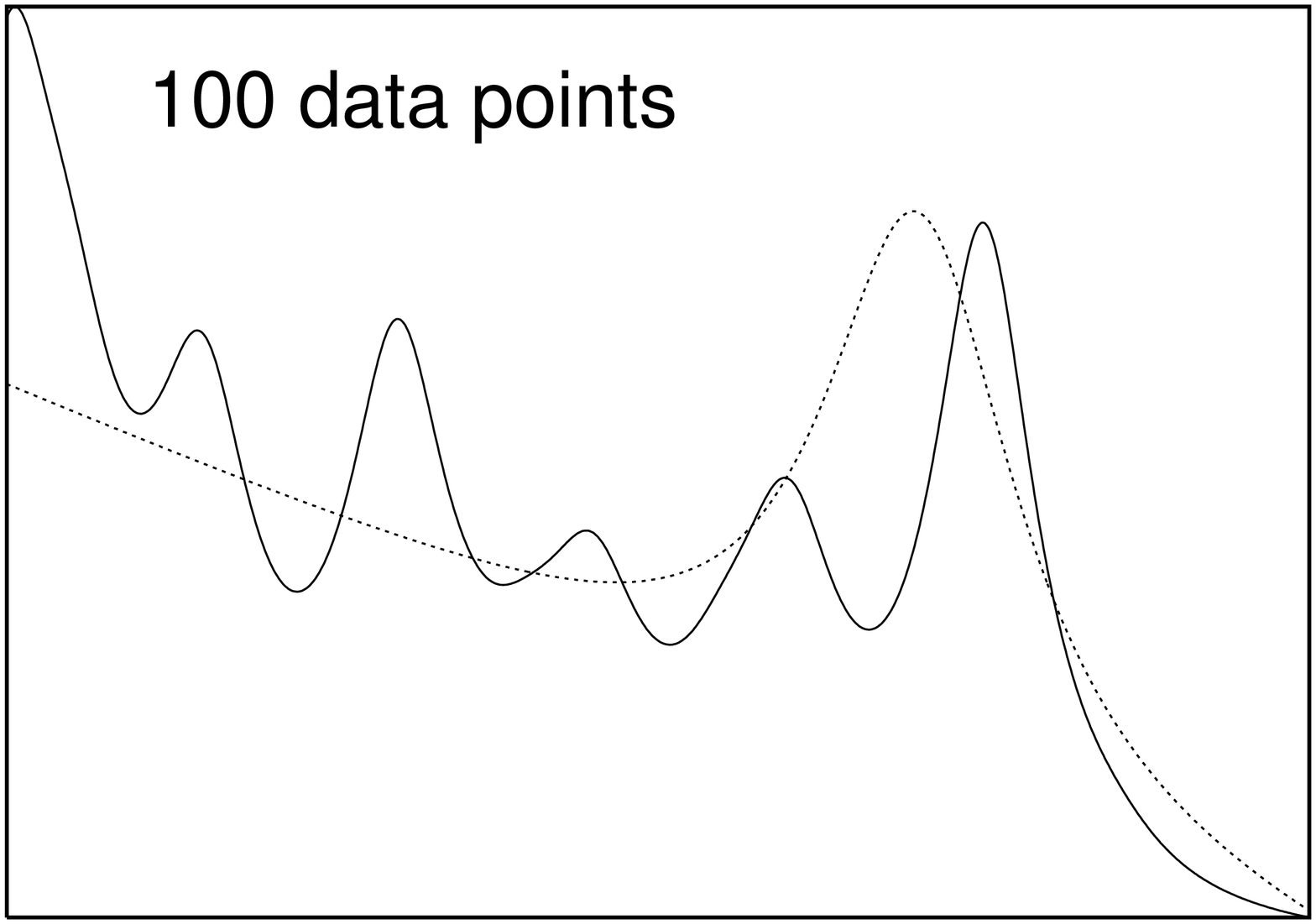,width=0.23\linewidth,angle=270}%
\epsfig{figure=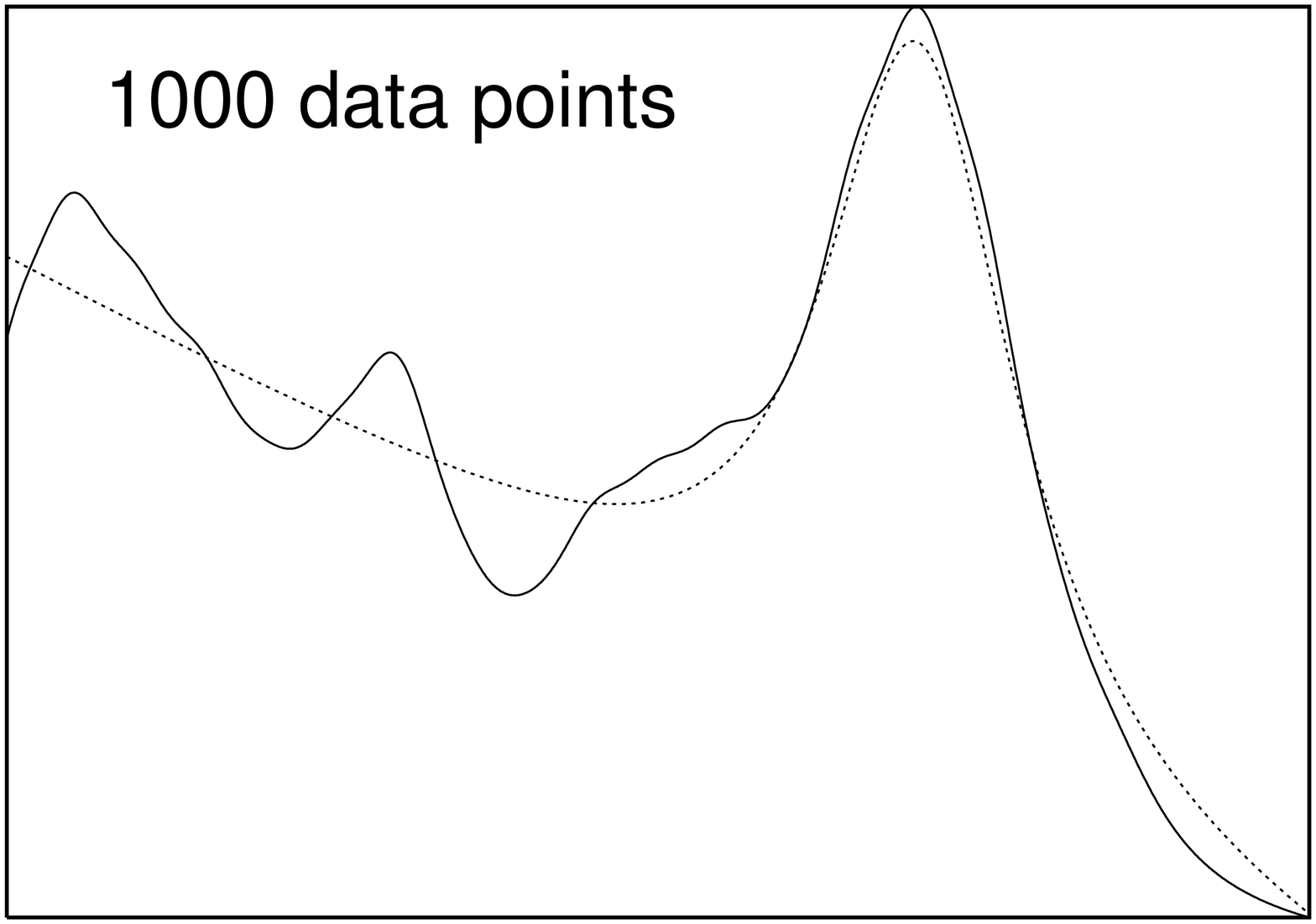,width=0.23\linewidth,angle=270}%
\epsfig{figure=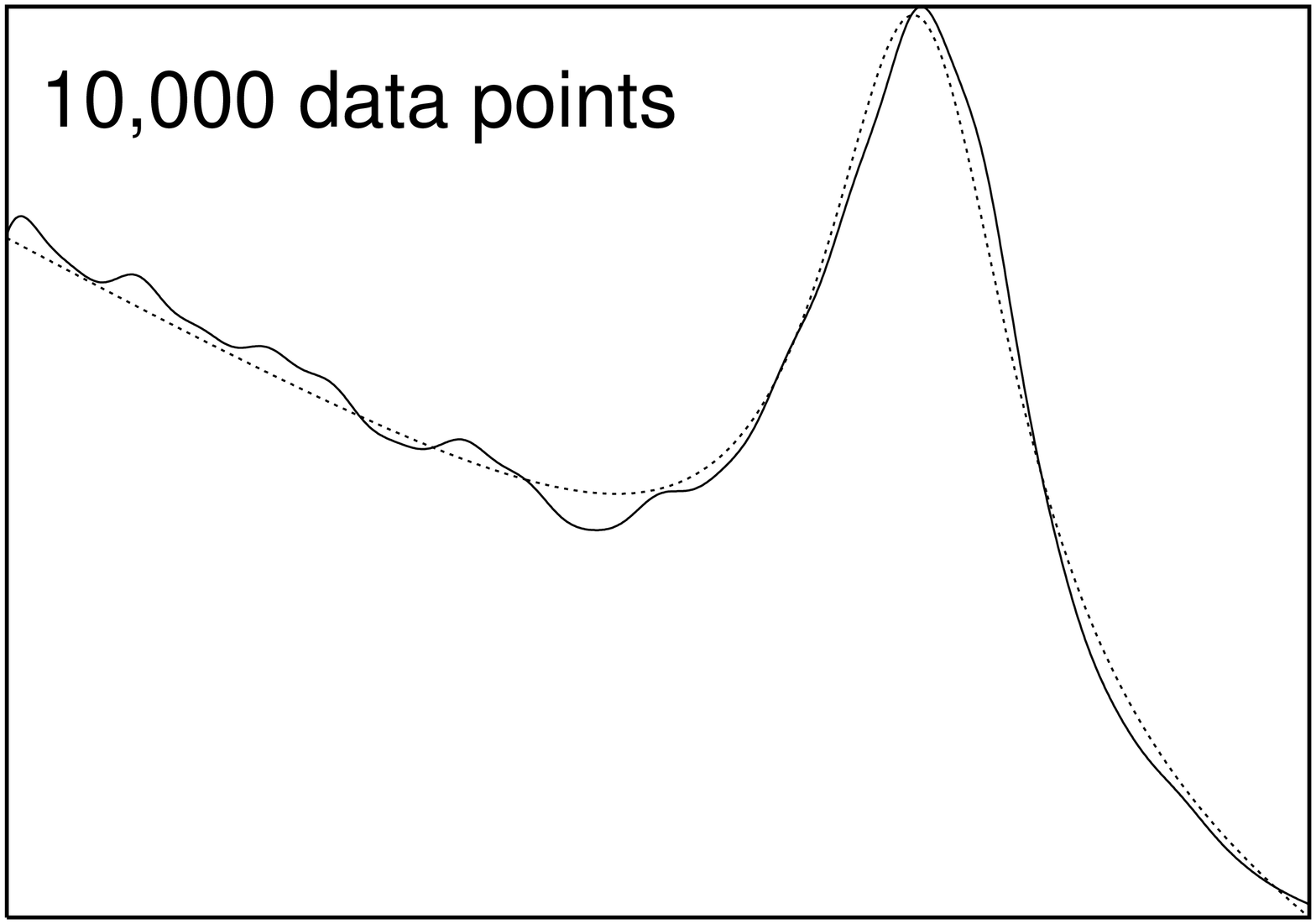,width=0.23\linewidth,angle=270}
\caption{\label{fig1}%
Histograms (upper graphs) and 
Cauchy-\upds\ (lower graphs) for random 
data-points, distributed following the dashed curve.}
\end{center}
\vspace{-10pt}
\end{figure}

\begin{figure}
\begin{center}
\epsfig{figure=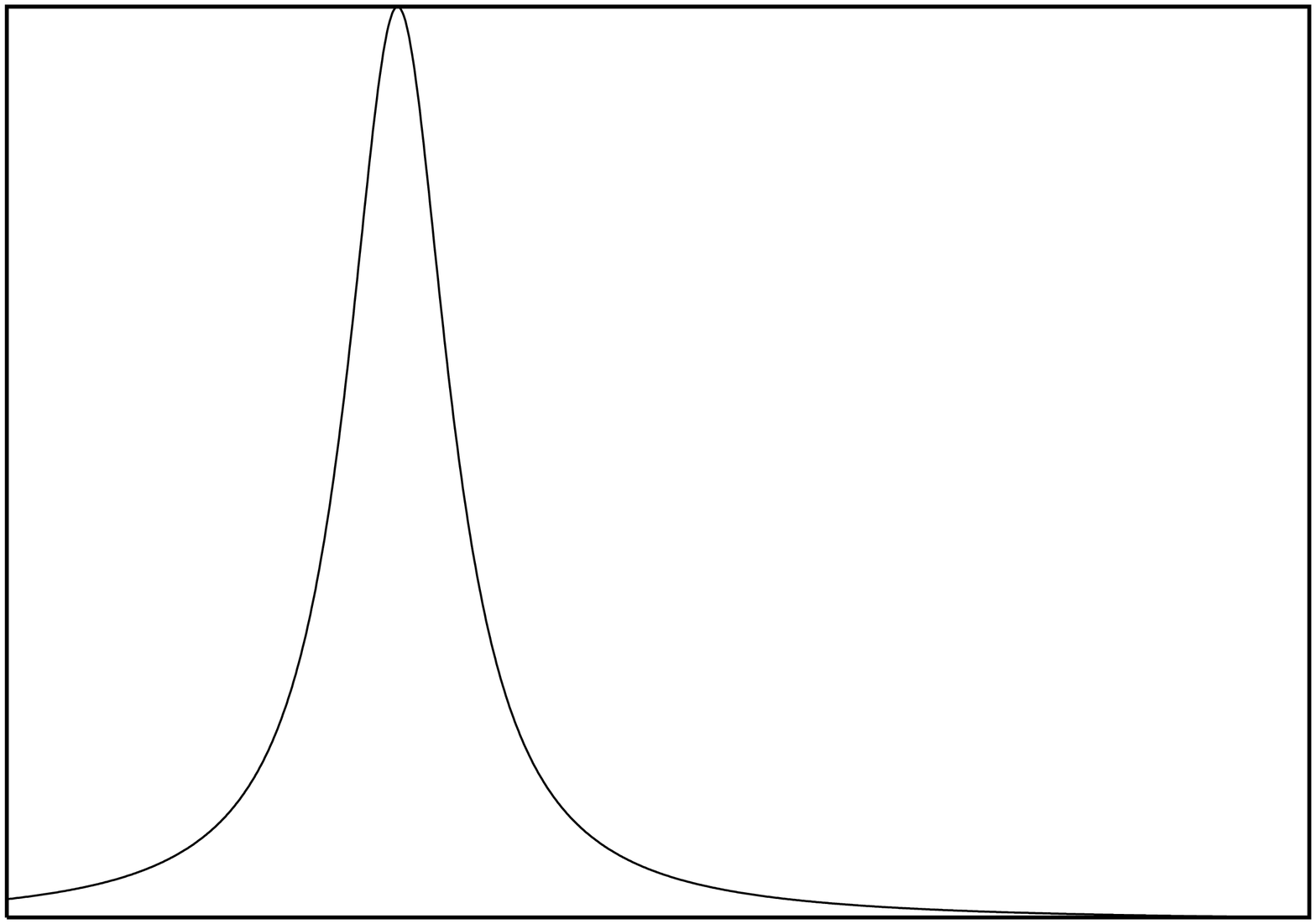,width=0.23\linewidth,angle=270}\hspace{30pt}
\epsfig{figure=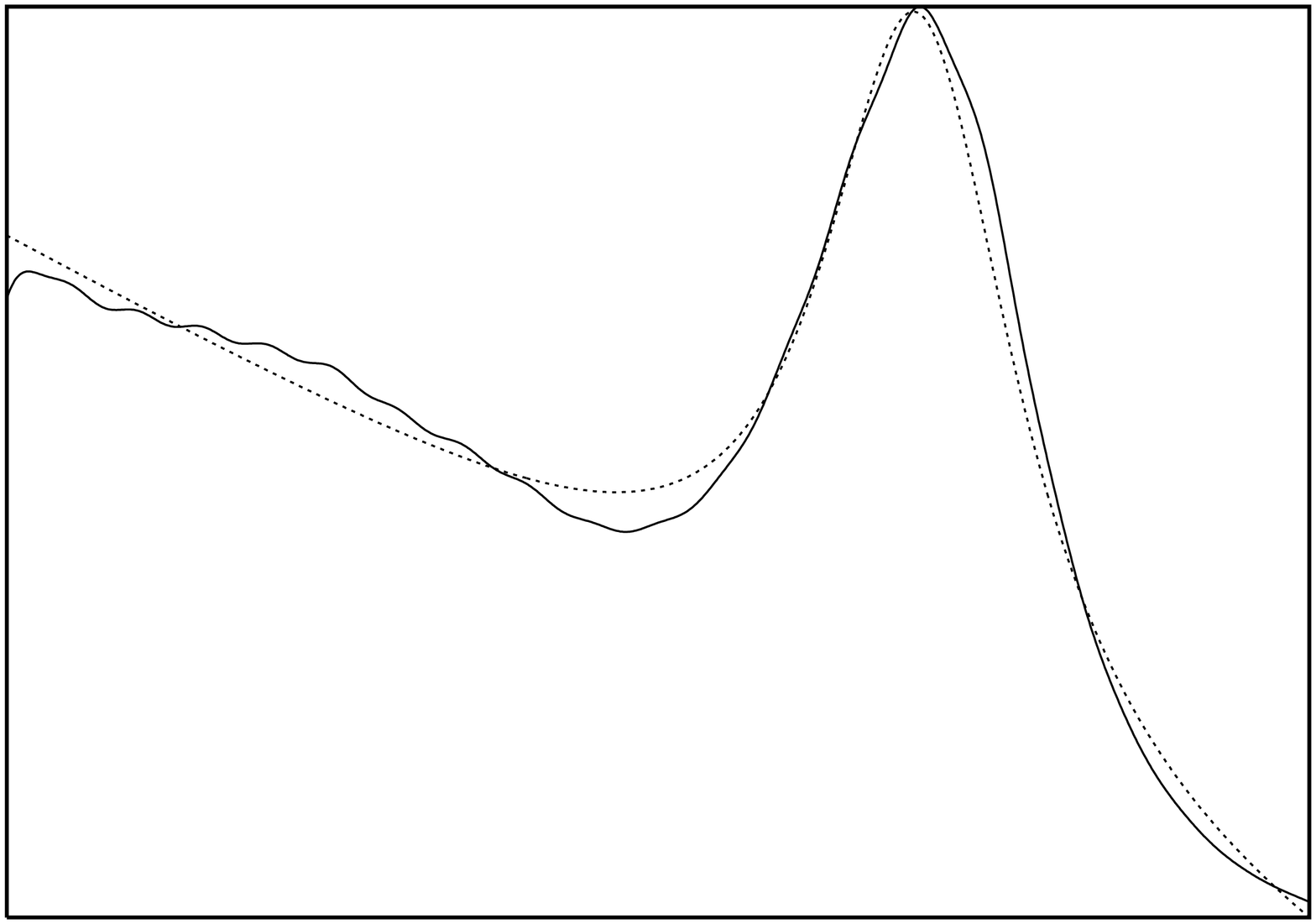,width=0.23\linewidth,angle=270}
\caption{\label{fig2}%
Left: $\gdns_7$ for $\nwe=21$. Right: $10^4$ data points taken in batches of
less than $10^3$, with only one iteration of \Algorithm{alg1} for each batch.}
\end{center}
\vspace{-10pt}
\end{figure}

As a small application, we show how the above may be used to construct
\upds.
For simplicity, we consider the one-dimensional case of a histogram on
the interval $[0,1]$.
Instead of 
$\nwe$
normalized indicator functions
$\gdns_i(\dap)=\nwe\theta(\dap-\frac{i-1}{\nwe})\theta(\frac{i}{\nwe}-\dap)$
of 
$\nwe$
bins with width
$1/\nwe$,
we use
$\nwe$
Cauchy densities 
\begin{displaymath}
   \gdns_i(\dap) \df 
   \frac{A_i}{1+(\srac{\dap-\dap_i}{\stdev_i})^2}
\quad\textrm{with}\quad
   \stdev_i = \srac{1}{\nwe} 
\;\;,\;\; 
   \dap_i = \srac{i-1}{\nwe-1}
\;\;,
\end{displaymath}
and normalization
$1/A_i=\arctan(\srac{1-\dap_i}{\stdev_i})+\arctan(\srac{\dap_i}{\stdev_i})$.
The left of \Figure{fig2} depicts $\gdns_7$ for $\nwe=21$.
In \Figure{fig1} we present the histograms (upper graphs) and Cauchy-\upds\ 
(lower graphs) for $10^2$, $10^3$ and $10^4$ random 
data-points, distributed following the density depicted with the dashed curve
in all graphs. 
Both types consist of $21$ channels, and for the generalized type the 
value of the weights after a maximum of $10^3$ iterations of \Algorithm{alg1}
is used.

For high number of data points, a high number of iterations becomes
inappropriate, and in case of a data stream, the number of data points
increases continuously.
In those cases, batches of data points can be used, and \Algorithm{alg1} can be
applied once with each batch.
In order to choose a size for the batches, we take into account the common rule
that, in a normal histogram, every bin should contain at least a few
data points in order to trust it.
The number of data points from a sample
$\Dap=(\dap_1,\dap_2,\ldots,\dap_{\nda})$
in the bin corresponding to indicator function
$\ind_i=\vol_i\gdns_i$
is given by
\begin{displaymath}
   \sum_{k=1}^{\nda}\ind_i(\dap_k)
   = \nda\vol_i\Omeas{\gdns_i}
   = \nda\frac{\Omeas{\gdns_i}}{\Pmeas{\gdns_i^2/\Pdns}}
\;\;,
\end{displaymath}
and we can use this for the general case. Of course,
$\Pmeas{\gdns_i^2/\Pdns}$
can only be estimated, for example with
$\Omeas{\gdns_i^2/\Gdns{\we}}$.
The right of \Figure{fig2} shows the result with $10^4$ random data points, 
taken in batches.
The size of the batches was such that the `generalized number of data points' 
for each channel was at least $35$, which happened to lead to batches with not
more than $10^3$ data points.

\section{Multi-channeling with adaptive channels\label{sec3}}
In the discussion so far, the channels
$\gdns_i$
were fixed, and the only adaptation appeared for the weights
$\we_i$.
As described in the introduction, it would be attractive to also adapt the 
channels.
We introduce the algorithm \parni
\footnote{{\tt P}ractical {\tt A}daptive {\tt R}andom {\tt N}umber 
{\tt I}dealizer.}
in which this is achieved.
We consider the 
$\dimO$-dimensional hypercube
$\dspace=[0,1]^{\dimO}$.
The channels are all normalized indicator functions, which are, however,
not necessarily non-overlapping.
The subsets corresponding to the indicator functions will only 
be {\em boxes} of the type
$[a_1,b_1]\times[a_2,b_2]\times\cdots\times[a_{\dimO},b_{\dimO}]$.
\begin{algorithm}[\parni]
\begin{enumerate}\initstep
\additem We start with $2\dimO$ channels corresponding to all pairs of boxes
         obtained by dissecting $\dspace$ in two along the Cartesian directions.
\end{enumerate}
\end{algorithm}
In the case of the data stream, data points
are generated from an external source, and in the case of numerical integration,
they are generated from 
$\Gdns{\we}$,
the density constructed with the channels and the weights.
For completeness, we repeat the algorithm to generate the points in the case
of numerical integration:
\begin{enumerate}\addsteps
\additem choose a channel with a probability equal to the weight of the channel;
\additem generate a point in the box corresponding to the channel, uniformly 
         distributed.
\end{enumerate}
A batch of data is collected, and
\begin{enumerate}\addsteps
\additem depending on the task, \Algorithm{alg1} or
         \Algorithm{alg2} (\Appendix{VarOpt})
         is applied to optimize the weights. 
\end{enumerate}
\begin{enumerate}\addsteps
\additem Directly after an optimization step, the box with the highest weight
         is replaced by $2\dimO$ boxes, obtained in pairs by dissecting the 
         original box in two along the  Cartesian directions.
\label{addboxes}
\end{enumerate}
This step makes the algorithm fully self-adaptive.
Executed for the first time in two dimensions, this step could look as follows
\begin{displaymath}
  \raisebox{-20pt}{\epsfig{figure=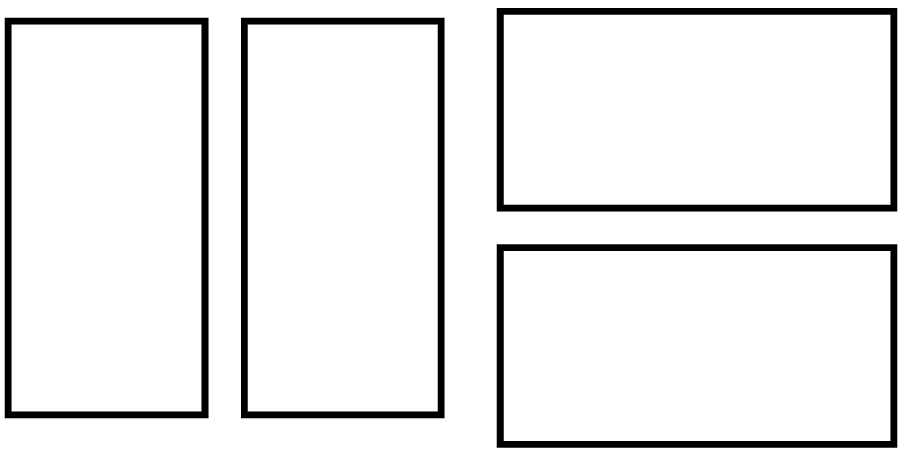,width=0.19\linewidth}}
  \quad\longleftarrow\quad
  \raisebox{-20pt}{\epsfig{figure=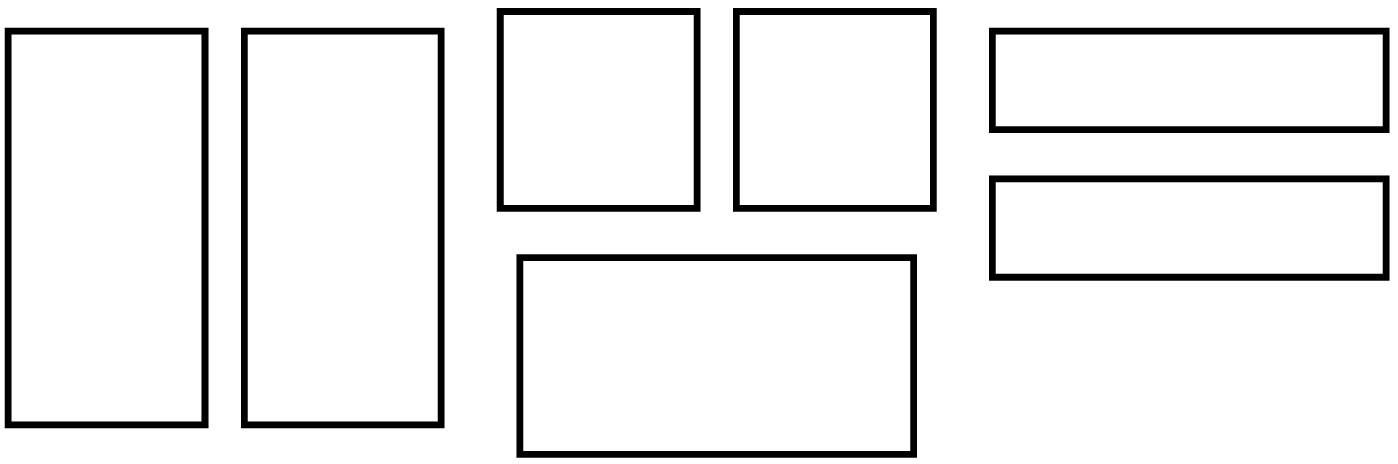,width=0.3\linewidth}}
\;\;.
\end{displaymath} 
The $4$ overlapping boxes on the l.h.s. cover the integration space twice. 
We just drew them in pairs next to each other for clarity.
Suppose the upper right box has the largest weight. Then it is replaced%
\footnote{Replacements or assignments are denoted with the arrow pointing to 
the left: $a\lar b$ means `$b$ is put in the memory space of $a$'.}
by 
the smaller boxes on the r.h.s., so that we end up with 
$7$ overlapping boxes which cover the integration space $2\frac{1}{2}$ times. 
The
addition of boxes cannot be continued forever in practice, 
and one would like to restrain the number of channels to a maximum.
\begin{enumerate}\addsteps
\additem If the number of channels reached its maximum, boxes with the 
         smallest weights can be merged by replacing them by the smallest
         possible (new) box that contains all of them.
\label{mergeboxes}
\end{enumerate}
Notice that this last procedure is very simple for Cartesian boxes. Also notice
that the merging of boxes is necessary, and that one should not just throw 
away boxes, because of the danger of ending up with `holes' in the integration 
space. 
Of course,
the last two steps can be repeated a few times before gathering new data points
in order to replace more channels at once and possibly accelerate the
optimization process.

One could ask the question why using all the overlapping boxes, and not use
just one new pair in each step. The answer to this question is simply stated by
a new question: which pair? The idea is to let the algorithm for the weight
optimization decide which new boxes are going to be important. Of course, the
number of new overlapping boxes in each step grows with the dimension of the
integration space, but not drasticly, only linearly. 

Another question could be why to use boxes in the first place, and why not some
other geometrical objects. The answer to this question is, firstly, the fact
that the complexity of the algorithm runs the risk of exploding with the
dimension of $\dspace$. For example, the mininal number of $\dimO$-simplices
needed to fill $[0,1]^{\dimO}$ is $\dimO!$, while the number of boxes needed is
$2$. Secondly, there is the practical simplicity to encode the boxes. 

\subsection{A simple application in two dimensions}
We present some results with probability density, or integrand, 
\begin{equation}
   \Pdns(\dap)
   \propto \frac{1}{(0.02)^2 + (\dap_{1}+\dap_{2}-1)^2}
\;\;,
\label{originaldistribution}\end{equation}
where the normalization is not written down explicitly for convenience. 
\begin{figure}
\begin{center}
\epsfig{figure=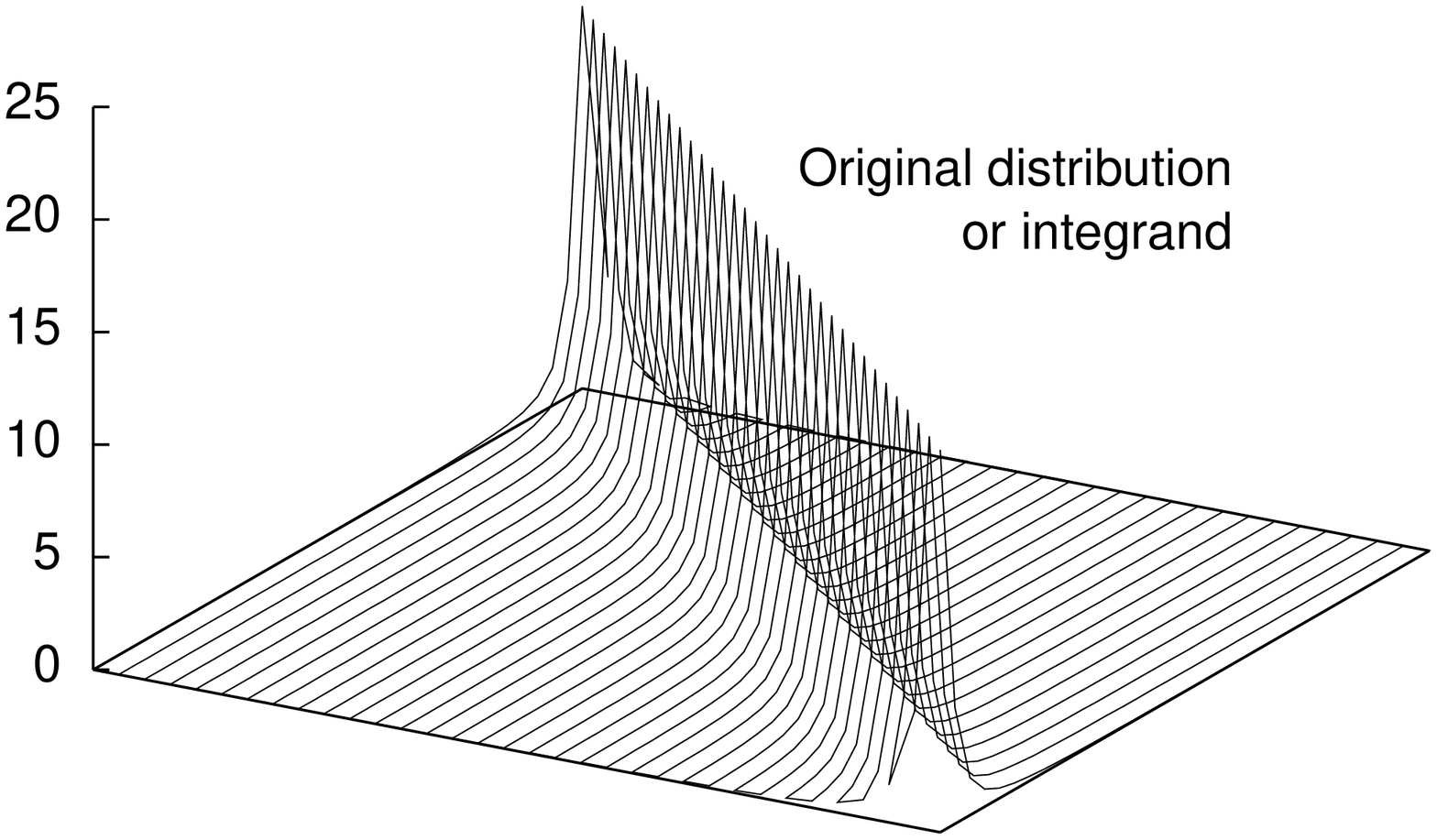,width=0.28\linewidth,angle=270}
\epsfig{figure=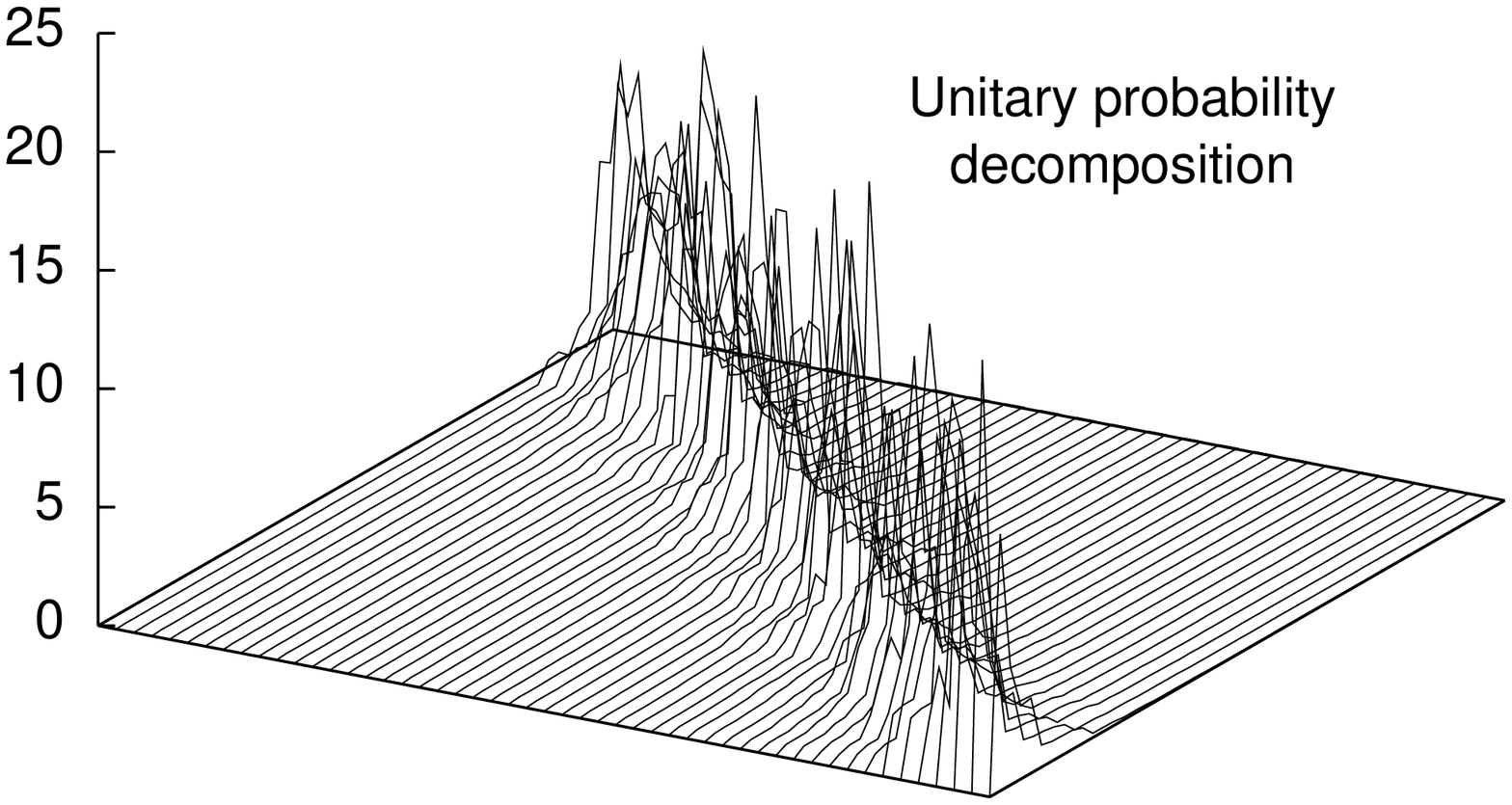,width=0.28\linewidth,angle=270}
\caption{\label{fig3}%
Left: $\Pdns(\dap)$ from (\ref{originaldistribution}).
Right: Unitary probability decomposition $\Gdns{\we}$ after $10^5$ 
data points distributed following $\Pdns$.}
\end{center}
\begin{center}
\begin{tabular}{|c||c|c|c||c|c|c|}
\hline
 \multirow{2}*{method} 
     & \multicolumn{3}{|c||}{all $10^5$ data points}
       & \multicolumn{3}{|c|}{last $10^3$ data points}  \\\cline{2-7}
   & result & error & eff. & result & error & eff.\\\hline\hline
  integration with variance optimization
   & $0.9977$ & $0.0031$ & $0.0156$ & $1.004$ & $0.014$ & $0.259$ \\\hline
  integration after \upd\ creation
   & $0.9980$ & $0.0016$ & $0.0865$ & $0.996$ & $0.016$ & $0.167$ \\\hline
  integration without optimization       
   & $1.0015$ & $0.0089$ & $0.0588$ & $1.093$ & $0.093$ & $0.059$ \\\hline
\end{tabular}
\end{center}
\begin{center}
Results for the integration of density (\ref{originaldistribution}).
\end{center}
\end{figure}
The choice for this particular density is taken from \cite{Jadach}.
It is interesting because it is not factorizable, and
{\tt VEGAS} performs badly when used to integrate it.
Since \parni\ uses the Cartesian boxes, also here 
this density constitutes a serious test.
All results with \parni\
were obtained with a maximal number of $1000$
channels, batches of $1000$ data points for the multi-channel optimization, 
and with $20$ iterations of steps \ref{addboxes} and \ref{mergeboxes}. 

The results for the integration of (\ref{originaldistribution}) are collected
in 
\Figure{fig3}.
Given are the result (average weight), the estimated error
(standard deviation) and the efficiency (average weight divided by maximum 
weight). 

For the method `integration with variance optimization', we see the effect of
the optimization if we compare the result of the last $10^3$ data points with
the result using all data points: the efficiency improves and the estimated
error is less than $10$ times worse, the `$10$' being expected from the
$1/\sqrt{\nda}$-rule of Monte Carlo integration.

For the method `integration after \upd\ creation', first $10^5$
data points distributed following (\ref{originaldistribution}) were generated to create the
\upd, and then $10^5$ integration points were generated using this \upd\ to integrate
(\ref{originaldistribution}).
So observing that the result is better than in the case of variance
optimization, one should keep in mind that twice as many data points have been
used. Furthermore, this method of integration cannot be considered useful as
long as we do not specify how to generate the first $10^5$ data points. We will
adress this issue in \Section{sec4}. For now, the ``smallness'' of the error
estimate should be interpreted as a measure of the ``goodness'' of the \upd.

The results of integration without any optimization are also included for 
comparison. Notice that in that case, the efficiency including the total 
amount of data is actually better than in the case of variance optimization. 
Apparently, the optimization process for
\parni\
needs some trail-and-error in the beginning to end up at the results
for the last $1000$ data points, which are obviously better than for the
non-optimized case. This problem becomes more apparent in more-dimensional
applications and can be solved following the method of
`integration after \upd\ creation', as we will see in \Section{sec4}.

\subsection{Application to phase space integration}
In the following, some results from the application of \parni\
in 
phase space integration are presented. We consider the problem of calculating
\begin{equation}
   \int d\Phi_n(Q;m_1^2,\ldots,m_n^2;p_1,\ldots,p_n)
        \,\Ant(p_0,p_1,\ldots,p_n,p_{n+1})
\;\;,
\label{antennaint}\end{equation}
with
\begin{equation}
   d\Phi_n(Q;m_1^2,\ldots,m_n^2;p_1,\ldots,p_n)
   = \bigg(\prod_{i=1}^{n}d^4p_i\dirac(p_i^2-m_i^2)\theta(p_i^0)\bigg)
        \dirac\Big(Q-\sum_{i=1}^{n}p_i\Big)
\;\;,
\label{phasespace}\end{equation}
and integrand
\begin{equation}
   \Ant(p_0,p_1,\ldots,p_n,p_{n+1})
   = \frac{(2\pi)^{4-3n}\,\prod_{j>i=0}^{n+1}\theta((p_i\cdot p_j)-\scut)}
           {(p_0\cdot p_1)(p_1\cdot p_2)(p_2\cdot p_3)
            \cdots(p_{n}\cdot p_{n+1})(p_{n+1}\cdot p_0)}
\;\;,
\label{antenna}\end{equation}
where $(p_i\cdot p_j)$ denotes the Lorentz invariant scalar product, 
$p_0=(\srac{1}{2},0,0,\srac{1}{2})$ and 
$p_{n+1}=(\srac{1}{2},0,0,-\srac{1}{2})$.
The
problem in calculating this integral is the pole structure in the scalar
products of the integrand. Although the integrand is regularized by a cut-off
$\scut$ in these scalar products, it still has a peak structure that makes
convergence in a straightforward numerical calculation problematic. Integrals
with this type of singularity structures are typically found in the phase space
integration of QCD amplitudes, and they are called {\em antenna pole
structures} \cite{vanHamerenPapadopoulos,vanHamerenKleiss}.

The straightforward numerical (Monte Carlo) calculation would consist of the
generation of $n$ random momenta $p_i$ uniformly distributed in phase space,
the bounded $(3n-4)$-dimensional subspace of $\Real^{4n}$ encoded in $d\Phi$,
and the calculation of the average of the integrand. Each set of random momenta
is constructed from a set of random numbers between $0$ and $1$, and the idea
is now to let \parni\
deliver these numbers. For the construction of the
momenta, there are two algorithms on the market: \rambo, which uses the
so-called {\em democratic} approach \cite{rambo}, and the hierarchical
construction of momenta (\hicom). The disadvantage of \rambo\
for our application
is that is needs $3n$ instead of $3n-4$ random numbers per $n$ momenta, and we
want to keep the dimension as low as possible%
\footnote{The original code by R.~Kleiss uses $4n$ random numbers per $n$
momenta, but this can easily be reduced to $3n$ with little extra cost.}.
Furthermore, the freedom one has in the actual implementation of \hicom\
allows
for an algorithm that is completely equivalent with \rambo\
in the case of
massless momenta. This implementation is presented in \Appendix{hicom}.

\begin{figure}
\begin{center}
\epsfig{figure=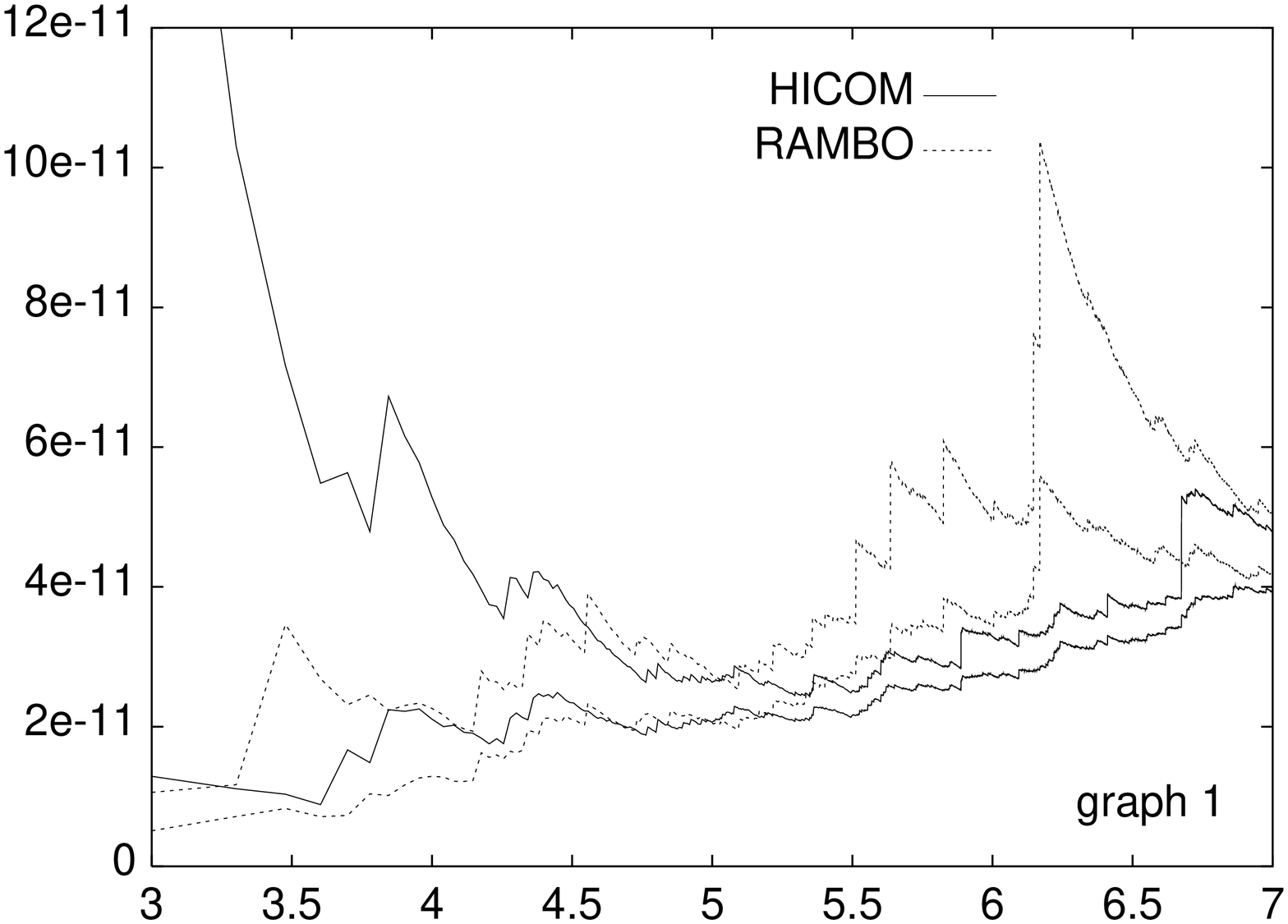,width=0.34\linewidth,angle=270}
\epsfig{figure=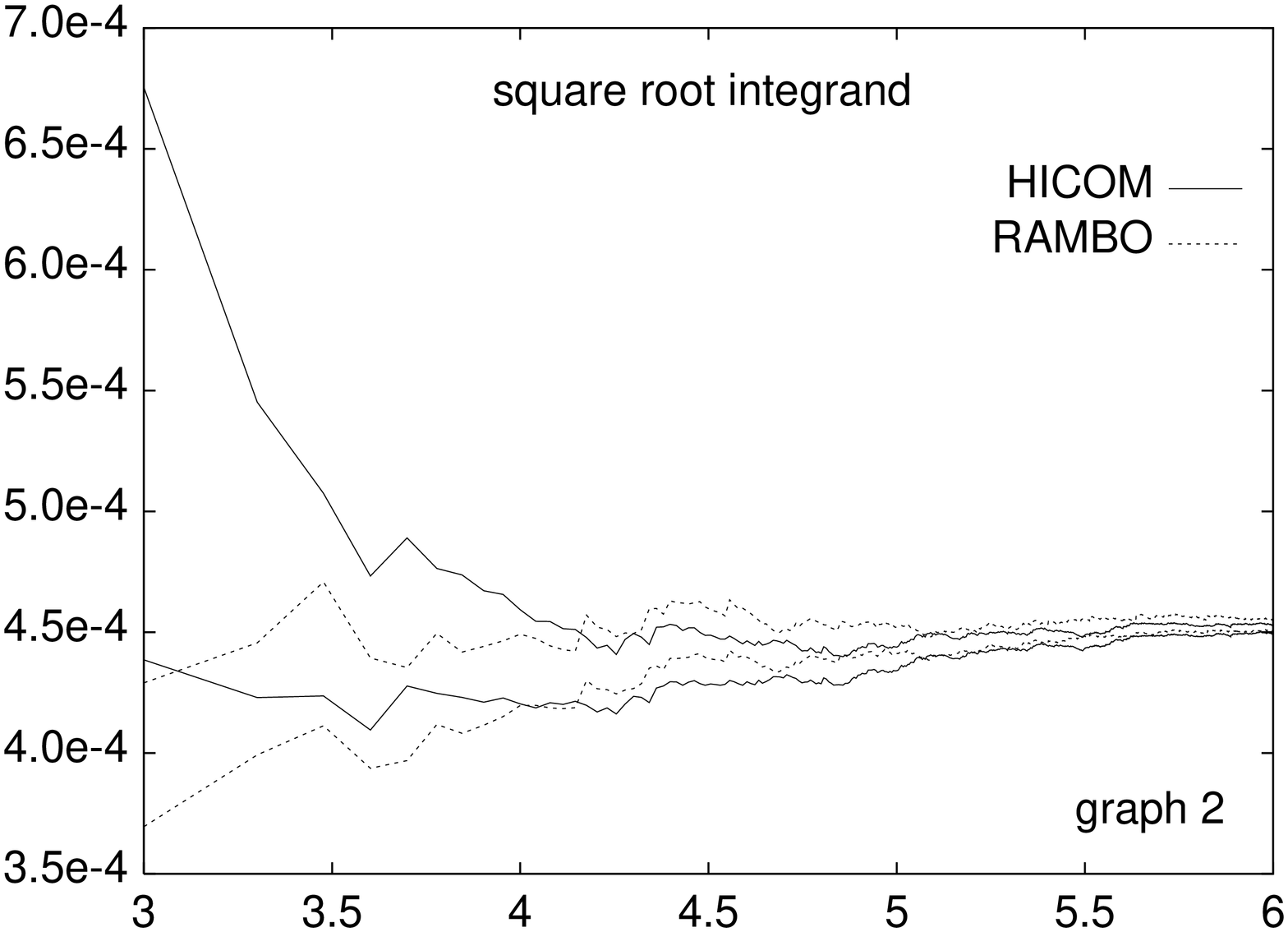,width=0.34\linewidth,angle=270}
\epsfig{figure=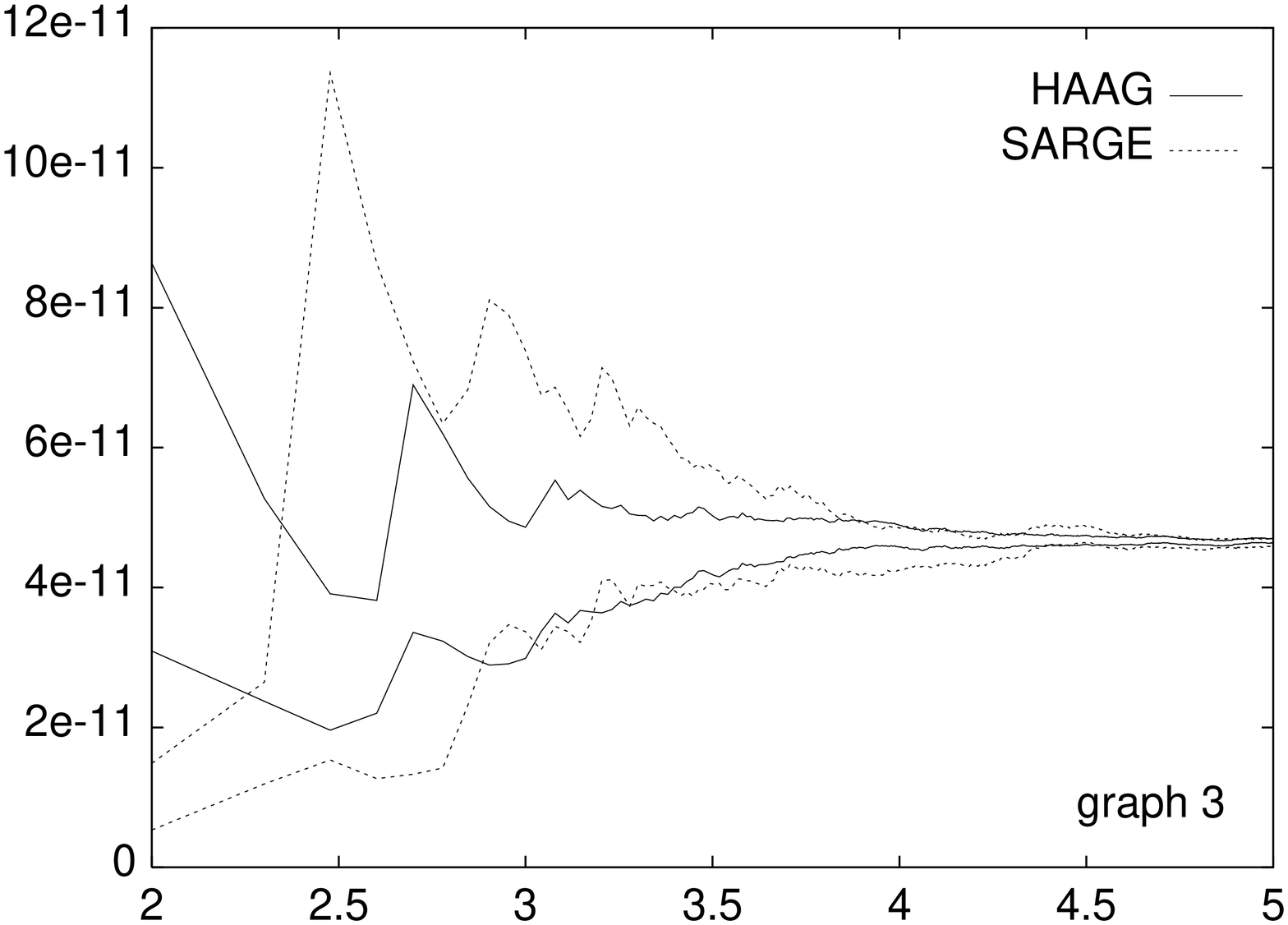,width=0.34\linewidth,angle=270}
\epsfig{figure=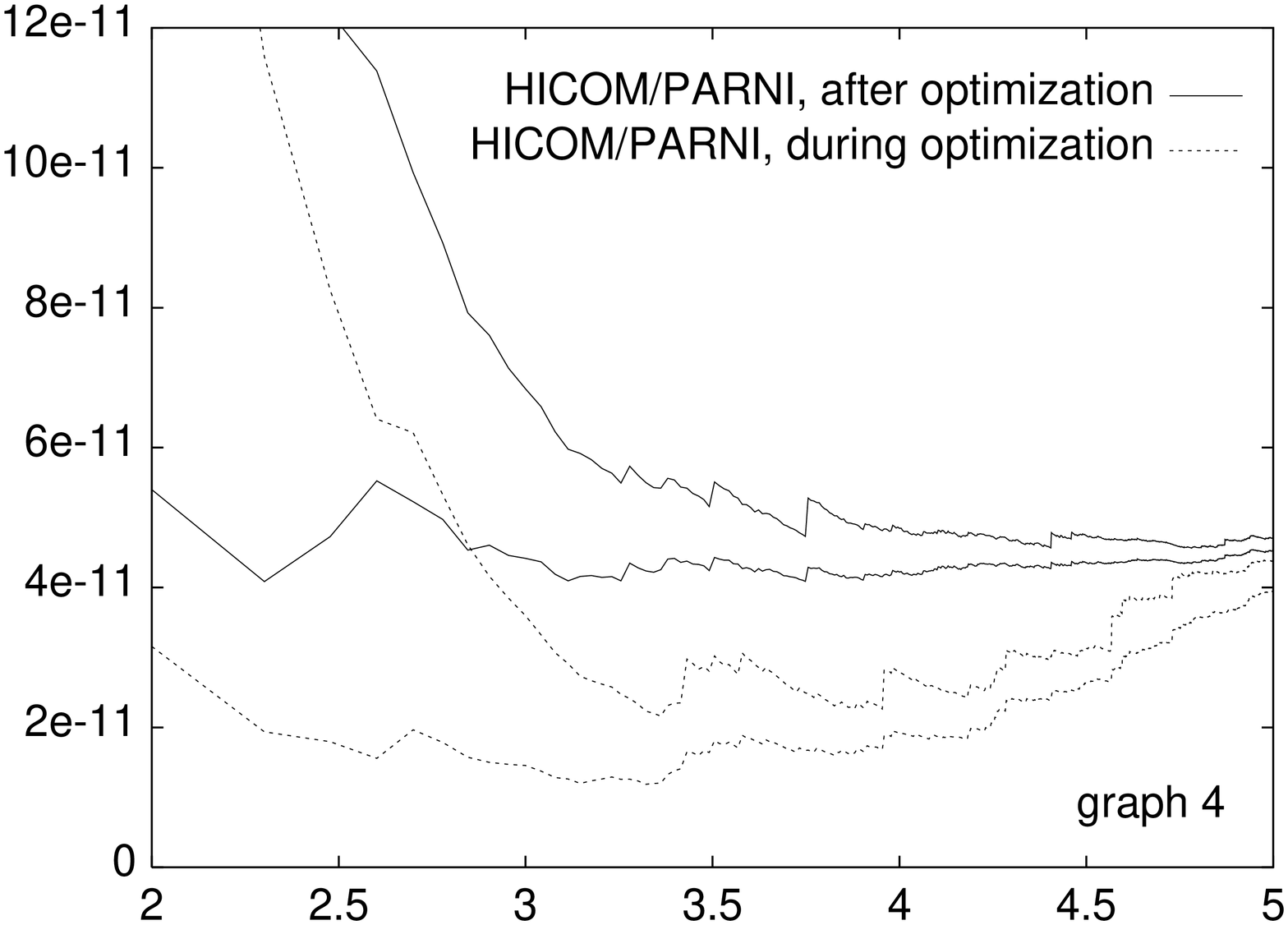,width=0.34\linewidth,angle=270}
\caption{\label{fig4}%
The process of convergence during the Monte Carlo calculation of
(\ref{antennaint}) for $n=4$ massless momenta with center of mass enery
$\sqrt{Q^2}=1000\,\mathrm{GeV}$ and $\scut=450\,\mathrm{GeV}^2$.  Along the
horizontal axis runs $^{10}\log(\textrm{\# events})$.  Two curves of the same
type give the average {\em plus} the standard deviation and the average {\em
minus} the standard
deviation.}
\end{center}
\end{figure}
\Figure{fig4} shows the results for the case of $n=4$ massless momenta
with a center of mass
energy $\sqrt{Q^2}=1000\,\mathrm{GeV}$ and a cut-off
$\scut=450\,\mathrm{GeV}^2$. 
There, the process of convergence during the Monte Carlo integration is shown.
Along the horizontal axis runs $^{10}\log \nda$, where $\nda$ is the number of
generated events (sets of momenta). In each graph, the two curves of the same
type give the average {\em plus} the standard deviation and the average {\em
minus} the standard deviation after the number of events on the horizontal
axis. 

Graph 3 shows the result with phase space generators \haag\ and \sarge, which
were specially designed to integrate functions with antenna pole structures
\cite{vanHamerenPapadopoulos,vanHamerenKleiss}. This is how a decent Monte
Carlo integration process should look like. The standard deviations converge to
zero and the average moves around a straight line: the process is unbiased and
the estimated error can be trusted at any point. 

\sarge\
and \haag\
were developed to cure graph 1, the result with \hicom\
and
\rambo. The standard deviations hardly converge and the estimated error cannot
be trusted at all: as long as no event hits a peak, the average is an
under estimation. If an event hits a peak, the standard deviation
increases drastically. Notice that the horizontal axis runs over $100$ times
more events than in graph 3.  Because of the particular peculiar behavior of
\hicom\
in this run, graph 2 with the result of the integration of the square
root of the integrand, which has a much softer singular behavior, is included.
We see a decent Monte Carlo process again and see that \hicom\
behaves the same
as \rambo.

Graph 4, finally, depicts the results with \hicom\
in combination with \parni.
For the curve `during optimization', \parni\
started from scratch. We see the
peculiar behavior of \hicom\
back in this run, for which \parni\
tries to
compensate. \parni\
used a maximal number of $1000$ channels, batches of
$1000$ data points for the multi-channel optimization with \Algorithm{alg2}, 
and $4$ iterations of
steps \ref{addboxes} and \ref{mergeboxes}.  After the generation of $10^5$
events, the channels of \parni\
were stored, and the process was repeated
starting with these channels, leading to the curves `after optimization', which
show a decent Monte Carlo process again.

\section{The use of \upds\ in numerical integration\label{sec4}}
The behavior of \parni\ during optimization in \Figure{fig4} is 
what one would naturally expect from a
general-purpose self adaptive Monte Carlo integration systems.  A
general-purpose system starts with no information about the integrand, and
cannot do anything else then start with generating integration points
distributed uniformly over the integration space. If the integrand has sharp
peaks, there is no enhanced probability for the system to hit these peaks
in the beginning, and an under estimation of the integral is the result. One can
imagine the extreme case of an integrand consisting of a sum of delta-peaks,
for which the probability to see them is zero.

A solution to this problem is given by the possibility to create a \upd\ before
starting with the integration. This \upd\ can then serve as a starting point
for the self adaptive integration system. 
In order to create the \upd, one needs a stream of data points that are
distributed following a density that looks like the integrand. This stream of
data will immediately show the peak structure, and in the extreme case
mentioned before it will {\em only} show the delta-peaks. Important in this
idea is that we realize that the creation of such a stream is, in principle, an
easier problem than the integration problem: one can, for example, use the 
Metropolis algorithm.

Since we want to restrict ourselves to integration problems in phase space like
in the previous section, we refer to \cite{Kharraziha} for the use of the
Metropolis algorithm.  The disadvantage of this algorithm for the use of
integration is that, although it creates a stream of data points that are
distributed following any normalized positive function on the integration
space, it does not give this normalization, the determination of which usually
is the actual integration problem one wants to solve. 

In the application we just described this normalization is not needed.
Having in mind, however, that the evaluation of the integrand is expensive, we
want to use all these evaluated integration points as efficiently as possible.
Let us denote
\begin{displaymath}
   \meas{f}_{\Pdns^{}} \df \frac{\int_{\dspace}f(\dap)\,\Pdns(\dap)d\dap}
                                {\meas{\Pdns}}
   \quad,\quad
   \meas{\Pdns} \df \int_{\dspace}\Pdns(\dap)d\dap
\end{displaymath}
for any integrable positive bounded function $\Pdns$ on the integration space
$\dspace$. With the Metropolis algorithm, one can estimate
$\meas{f}_{\Pdns^{}}$ for any quadratically integrable $f$, but one cannot
calculate $\meas{\Pdns}$. 
Inspired by the solution to this problem proposed in \cite{Kharraziha}, we 
observe that
\begin{displaymath}
   \meas{\Pdns}
   = \meas{\Pdns^{1-\alpha}}_{\Pdns^\alpha}\meas{\Pdns^\alpha}
\end{displaymath}
for any $\alpha\in[0,1]$.  We also observe that the function $\Pdns^\alpha$
will be easier to integrate than $\Pdns$ itself because the peaks are
suppressed. There is an equilibrium: for small $\alpha$ the calculation of
$\meas{\Pdns^\alpha}$ will be easy, but the estimation of
$\meas{\Pdns^{1-\alpha}}_{\Pdns^\alpha}$ will be difficult, whereas for
$\alpha$ close to $1$ the estimation of
$\meas{\Pdns^{1-\alpha}}_{\Pdns^\alpha}$ will be easy, but the estimation of
$\meas{\Pdns^\alpha}$ difficult. Important in light of the foregoing
is that {\sl the estimation of $\meas{\Pdns^{1-\alpha}}_{\Pdns^\alpha}$ will
serve a stream of data points distributed following $\Pdns^\alpha$ which can be
used to create a \upd\ for the estimation of $\meas{\Pdns^\alpha}$}.
This sounds tricky, like one uses integration points twice, but this is not the
case. The integration points used to estimate
$\meas{\Pdns^{1-\alpha}}_{\Pdns^\alpha}$ are used to initialize an integration
system to calculate $\meas{\Pdns^\alpha}$ with new integration
points.

The question that remains is what value to choose for $\alpha$. We shall stick
to values close to $1$, so that the estimation of 
$\meas{\Pdns^{1-\alpha}}_{\Pdns^\alpha}$ 
is relatively easy, and the problem of calculating  
$\meas{\Pdns^\alpha}$ is close to the original integration problem.

\subsection{Application to phase space integration}
It appears that \parni\ performs better with a small change in the algorithm,
namely by merging a number of boxes after every optimization step, so by
performing step \ref{mergeboxes} even before the maximum allowed number of
channels has been reached. The positive effect must be a result of the gain in
flexibility.  For the following calculations, a maximum number of $4000$
channels has been used, with an optimization step after every $4000$ events.
During every such step, more-or-less $50$ boxes were merged and step
\ref{addboxes} was iterated as many times as it takes to add more-or-less $120$
boxes. 

Another advantage of the use of \hicom\ instead of \rambo\ in combination with
\parni\ shows up for the use of the hybrid Metropolis procedure sketched
before, namely the fact that the mapping from random points in the 
$(3n-4)$-dimensional hypercube
to momenta is
invertible. This means that, for the initialization of \parni\ through the
creation of the \upd, the momenta generated with the Metropolis algorithm as
described in \cite{Kharraziha} can be used. One could, of course, use
another implementation of the Metropolis algorithm in the $(3n-4)$-dimensional
hypercube and map these points to momenta, but we do not want to go into this
issue at this stage, not in the least place because of the (extreme) simplicity
of the algorithm from \cite{Kharraziha}.

\begin{figure}
\begin{center}
\epsfig{figure=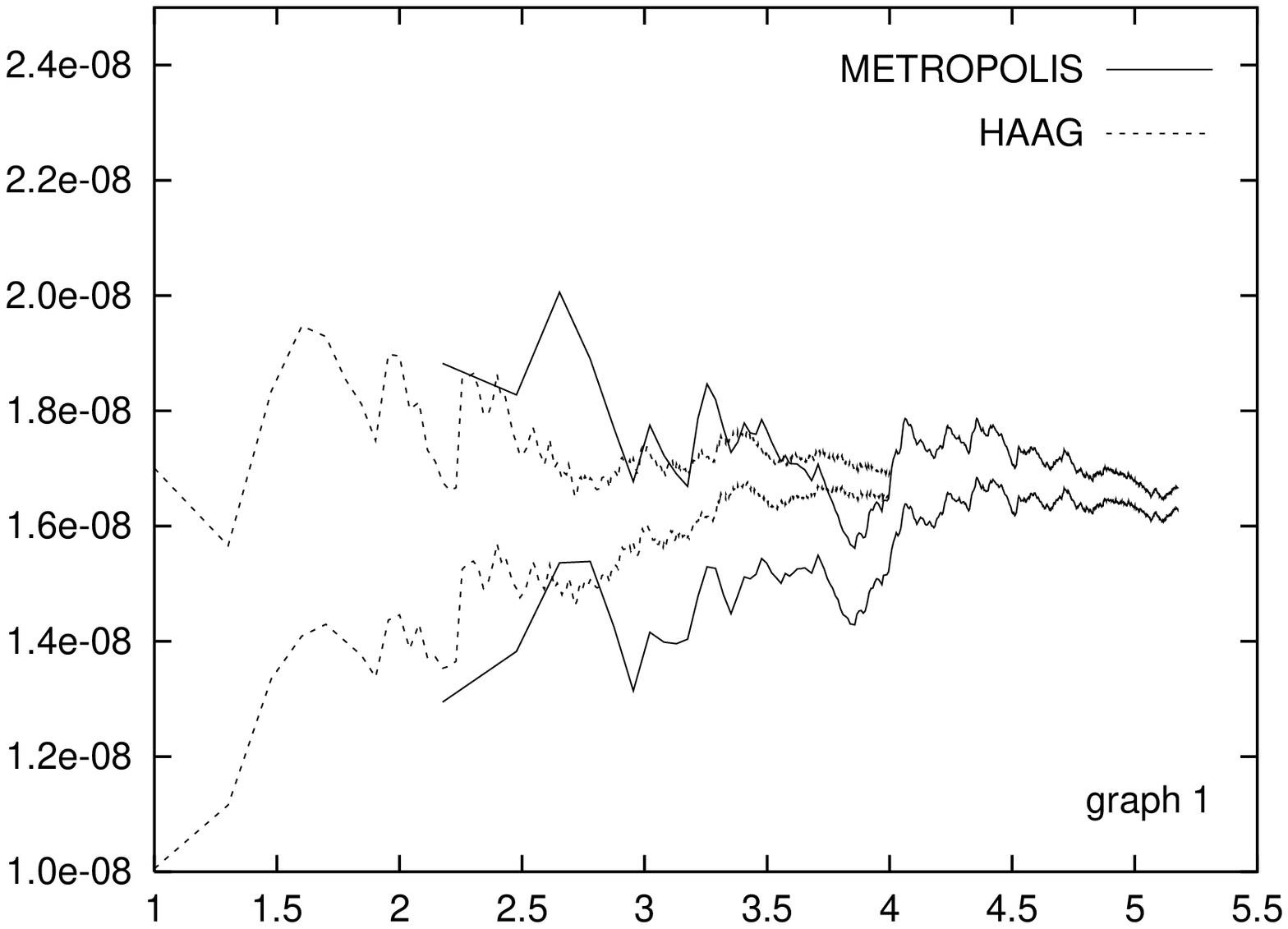,width=0.36\linewidth,angle=270}
\epsfig{figure=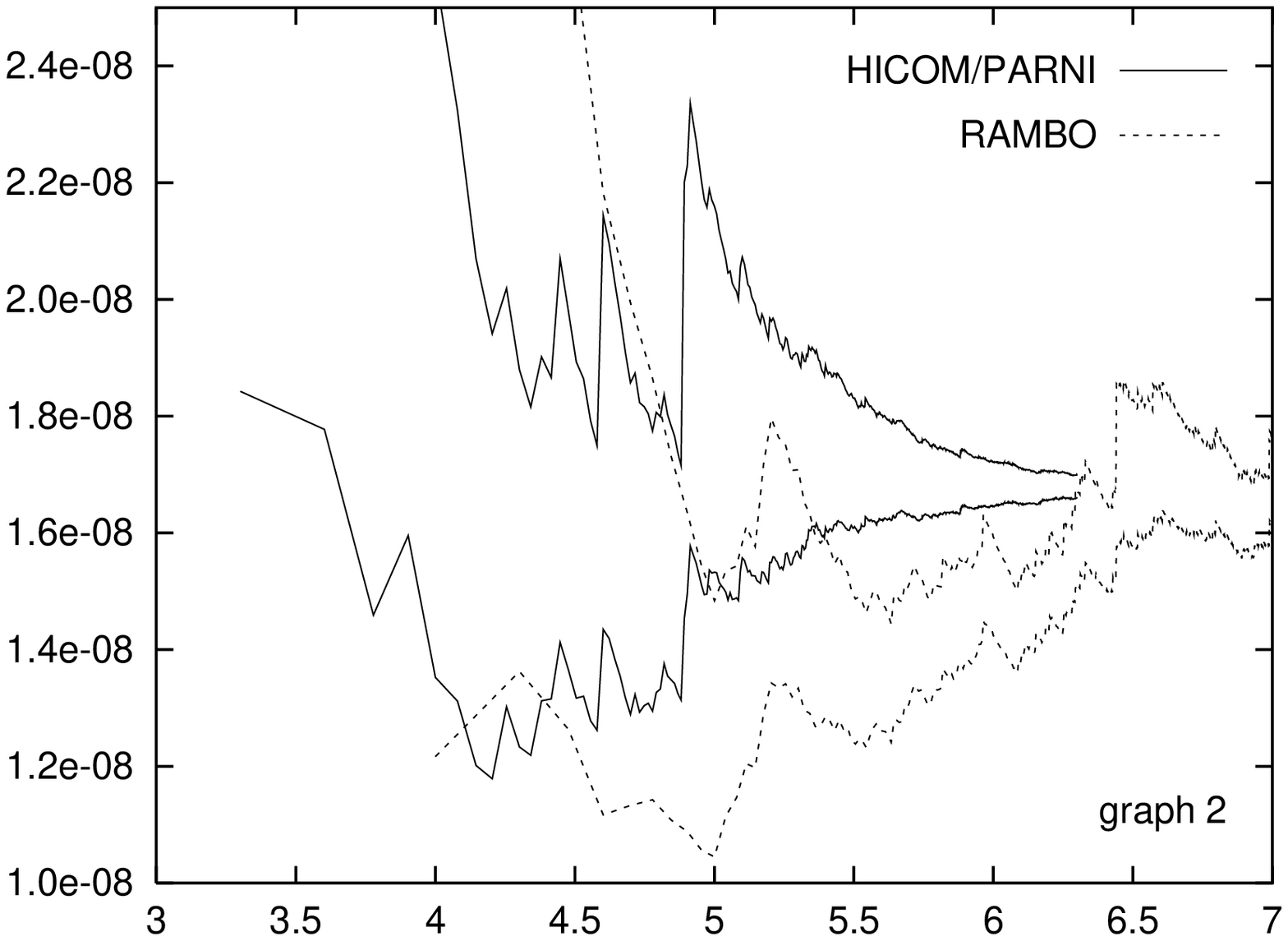,width=0.36\linewidth,angle=270}
\epsfig{figure=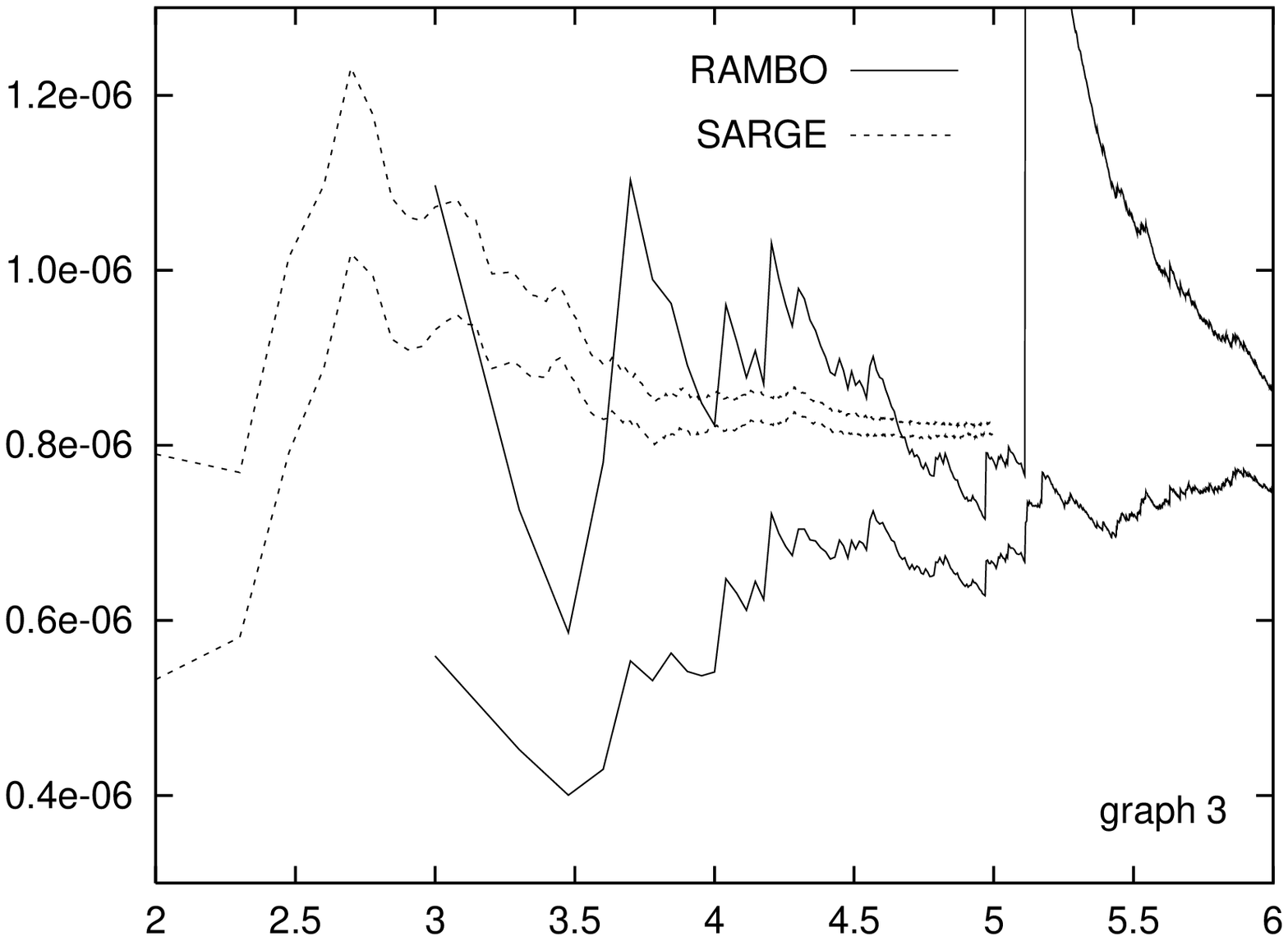,width=0.36\linewidth,angle=270}
\epsfig{figure=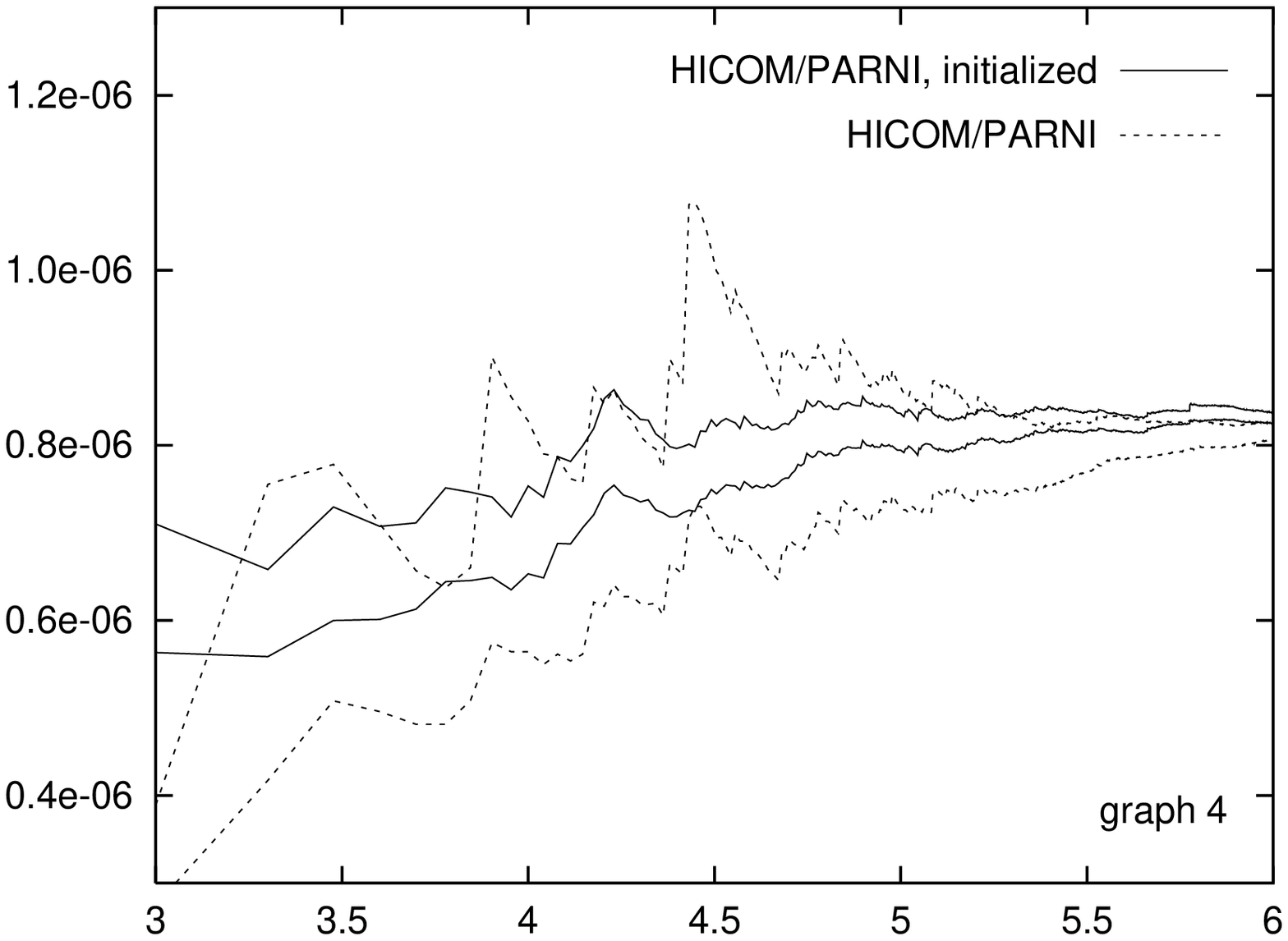,width=0.36\linewidth,angle=270}
\caption{\label{fig5}%
The process of convergence during the Monte Carlo calculation of
(\ref{antennasym}) for $n=4$ massless momenta with center of mass enery
$\sqrt{Q^2}=1000\,\mathrm{GeV}$ and
$\scut=450\,\mathrm{GeV}^2$. Along the horizontal axis runs
$^{10}\log(\textrm{\# events})$. Two curves of the same type give the average
{\em plus} the standard deviation and the average {\em minus} the standard
deviation.}
\end{center}
\begin{center}
\begin{tabular}{|c|c|c||c|c|c|c|c|}
\hline
  \multirow{2}*{integrand}
 &\multirow{2}*{graph}
 &\multirow{2}*{algorithm}
     & $I      $ & $\stdev$ & $\Nge  $ & $\stdev^2\Ttot/\Tint$ & $\stdev^2\Nac$ \\
 & & & $10^{-8}$ & $\%    $ & $10^{3}$ & $10^{3}             $ & $10^{3}      $ \\\hline\hline
  \multirow{4}*{$\Asym$} 
 &\multirow{2}*{1} &Metropolis& $1.649$ & $1.11$ & $150  $ & $0.0458$ & $0.0186 $ \\ \cline{3-8}
 &                 &     \haag& $1.669$ & $1.27$ & $10   $ & $0.0261$ & $0.00113$ \\ \cline{2-8}
 &\multirow{2}*{2} &    \parni& $1.670$ & $1.17$ & $2000 $ & $3.03  $ & $0.211  $ \\ \cline{3-8}
 &                 &    \rambo& $1.718$ & $4.70$ & $10000$ & $22.7  $ & $20.9   $ \\\hline\hline
  \multirow{4}*{$(\Asym)^{0.9}$} 
 &\multirow{2}*{3} &             \sarge& $81.89$ & $0.728$ & $100 $ & $0.00828$ & $0.00247$ \\\cline{3-8}
 &                 &             \rambo& $81.17$ & $7.11 $ & $1000$ & $5.20   $ & $4.79   $ \\\cline{2-8}
 &\multirow{2}*{4} &\parni, initialized& $83.16$ & $0.732$ & $1000$ & $0.587  $ & $0.043  $ \\\cline{3-8}
 &                 &             \parni& $81.66$ & $1.21 $ & $1000$ & $1.55   $ & $0.121  $ \\\hline
\end{tabular}
\end{center}
The final results corresponding with \Figure{fig5}. $I$ is the integral,
$\stdev$ the standard deviation, $\nda$ the number of generated events, 
$\Ttot$ the total computation time, $\Tint$ the time it takes to perform one
evaluation of the integrand and $\Nac$ the number of accepted (non-zero weight)
events.  
\end{figure}

\begin{figure}
\begin{center}
\epsfig{figure=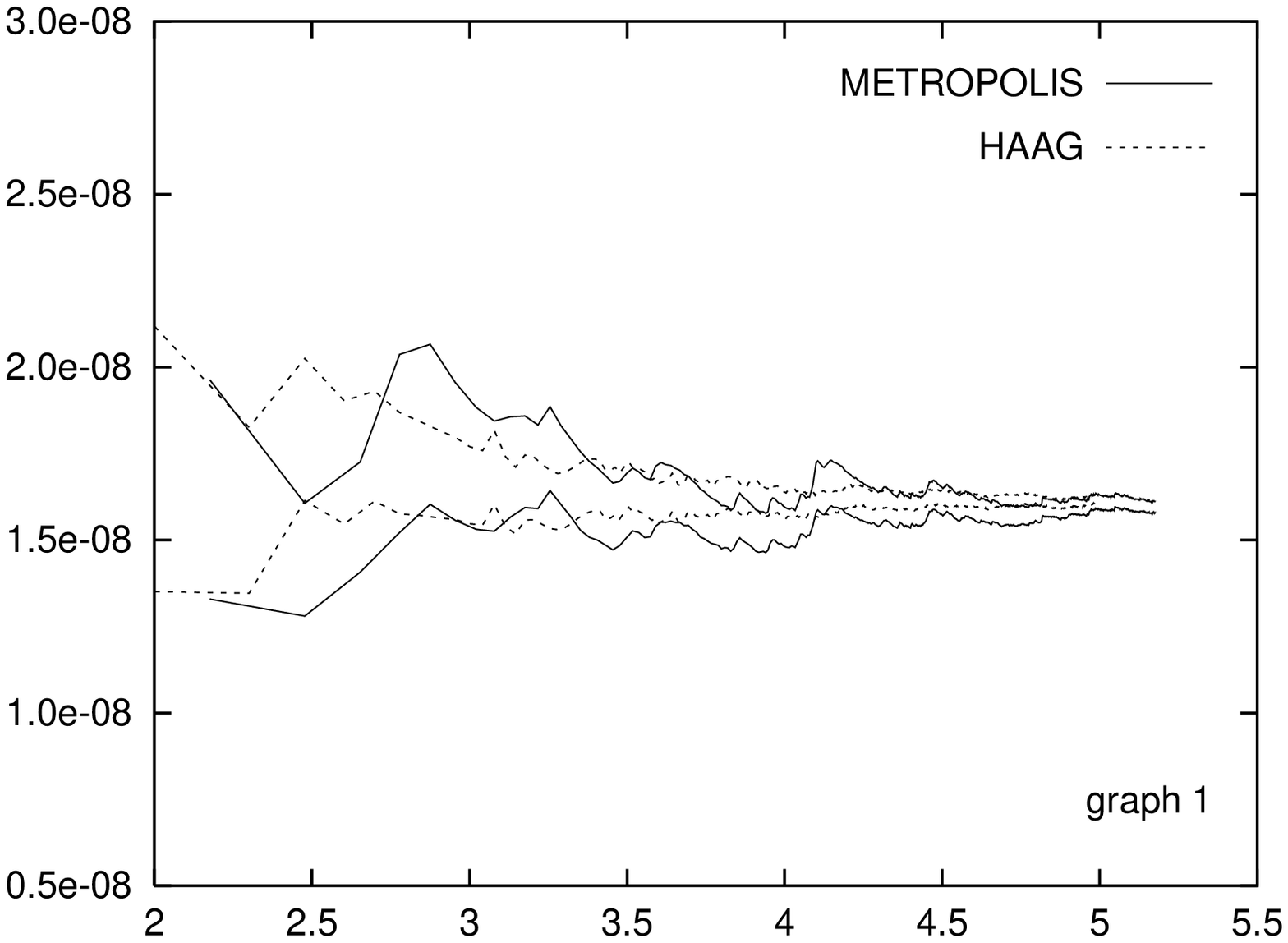,width=0.36\linewidth,angle=270}
\epsfig{figure=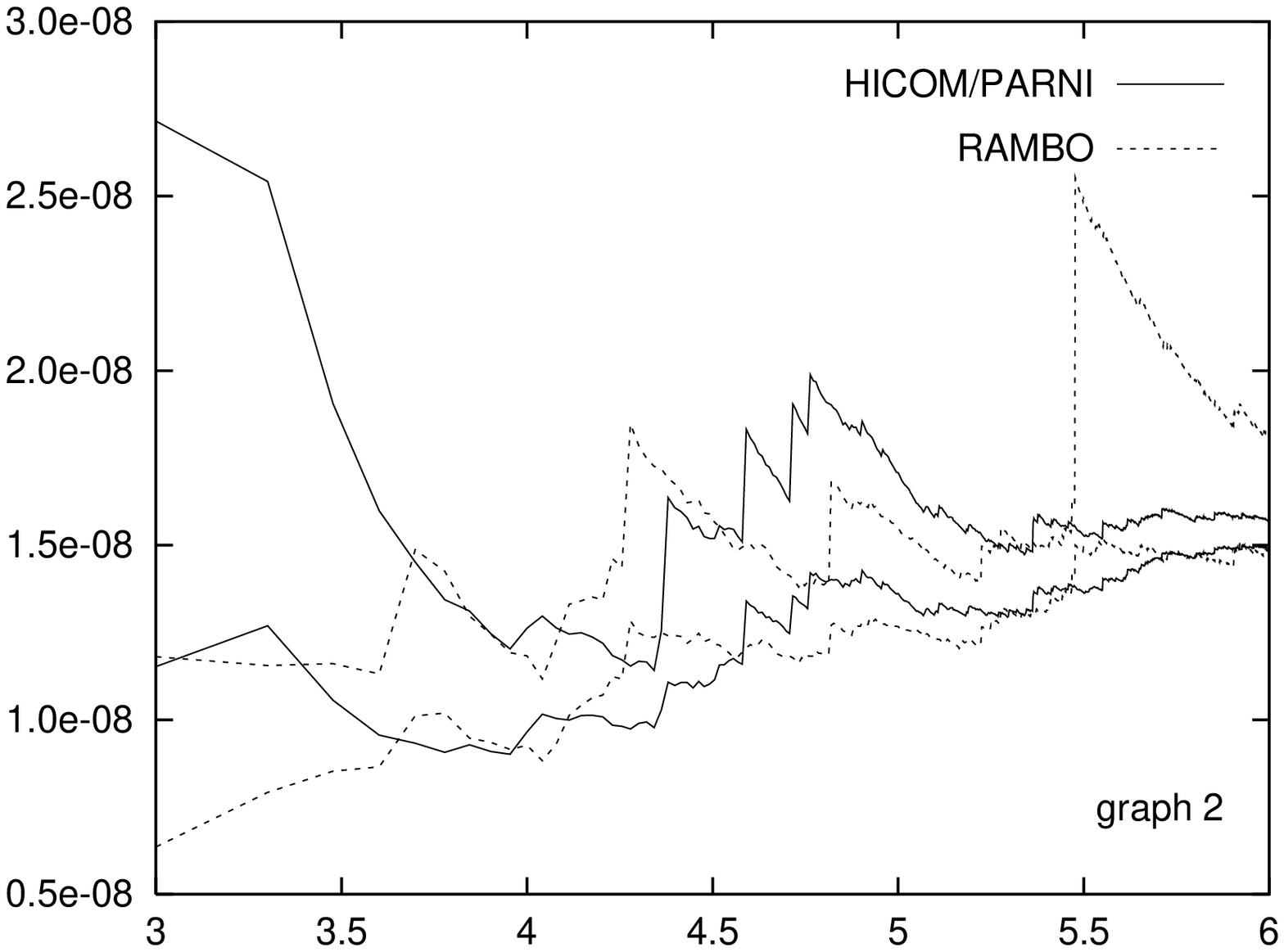,width=0.36\linewidth,angle=270}
\epsfig{figure=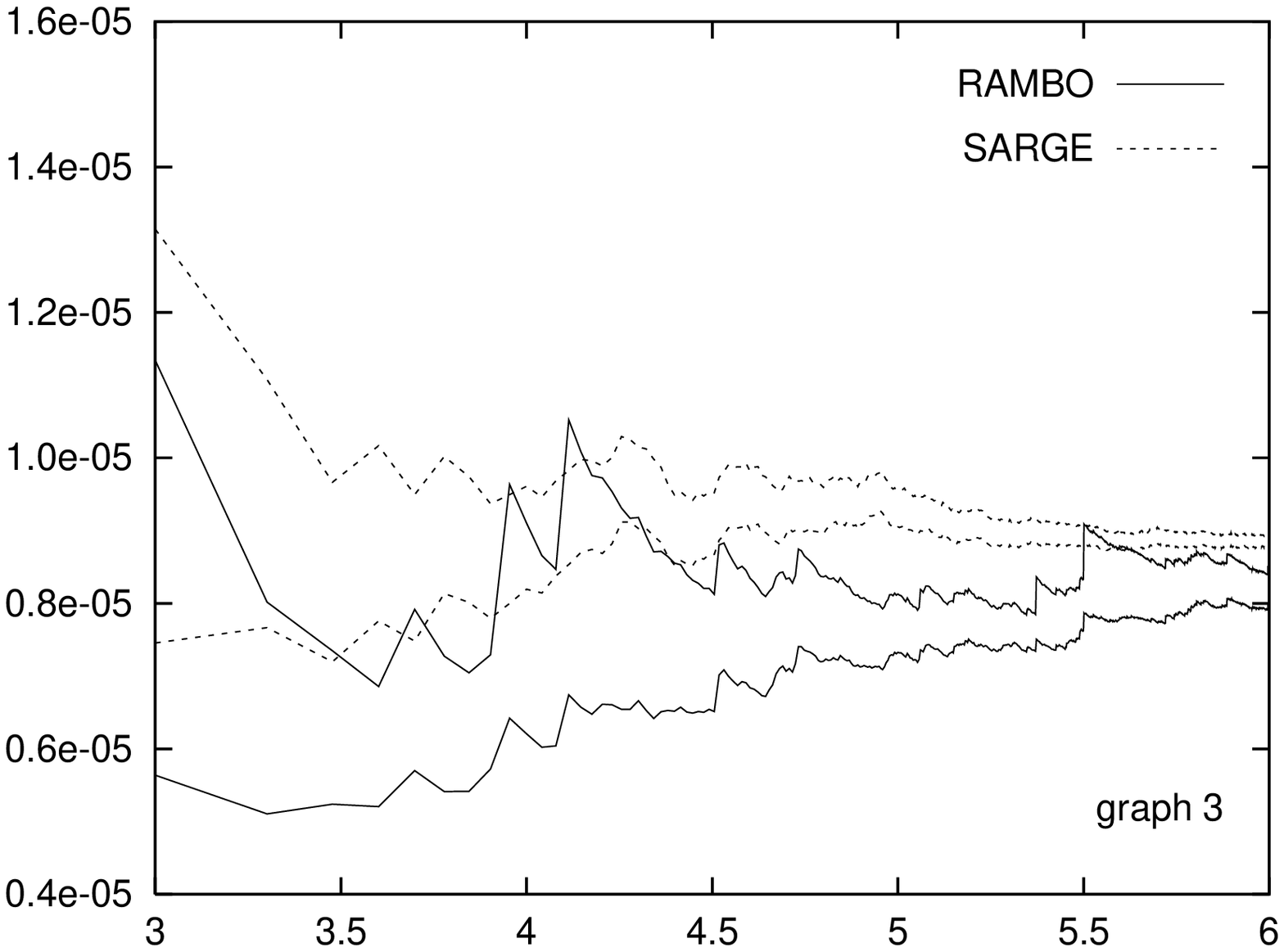,width=0.36\linewidth,angle=270}
\epsfig{figure=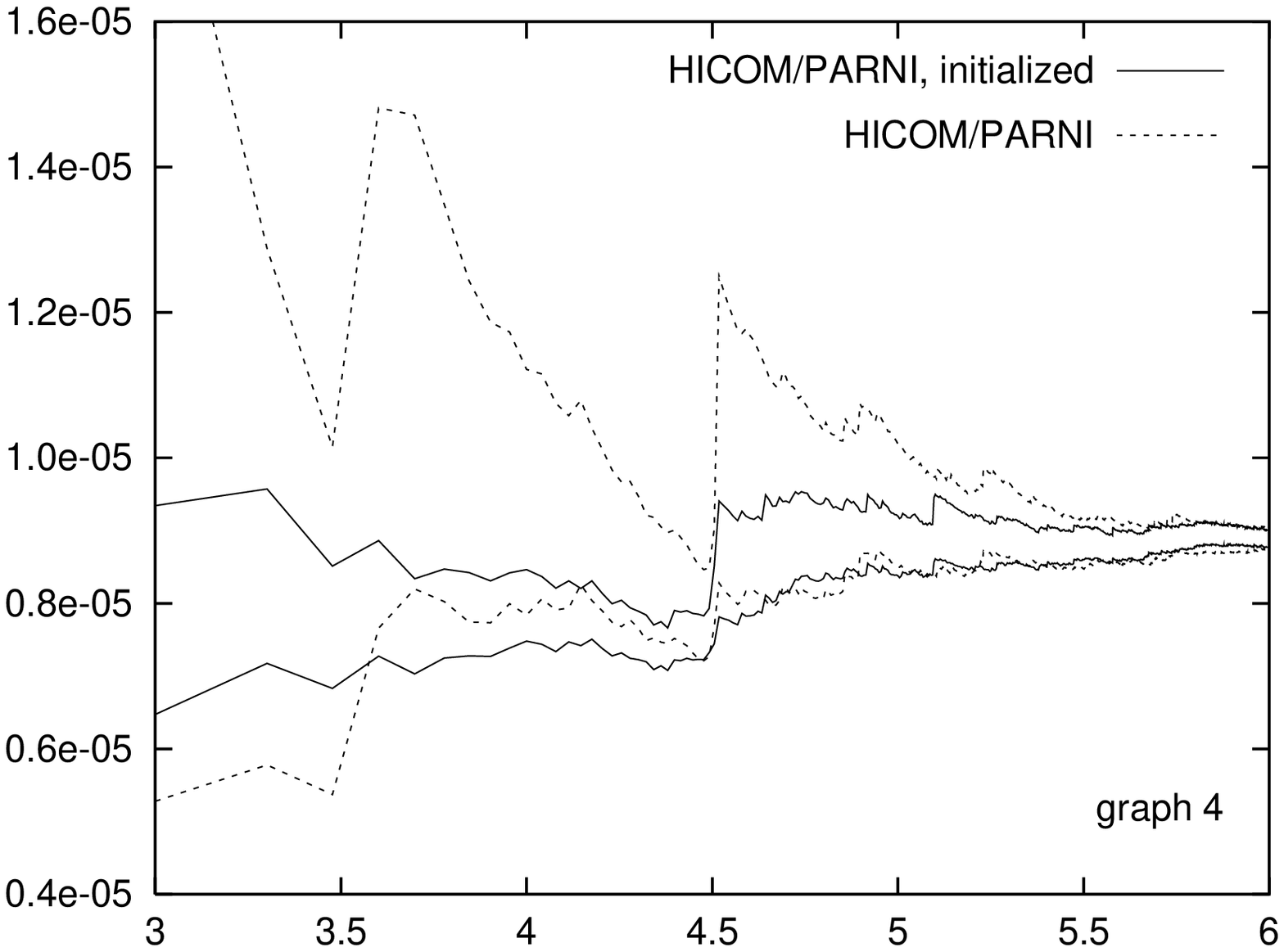,width=0.36\linewidth,angle=270}
\caption{\label{fig6}%
The process of convergence during the Monte Carlo calculation of
(\ref{antennasym}) for $n=6$ massless momenta with center of mass energy
$\sqrt{Q^2}=1000\,\mathrm{GeV}$ and
$\scut=450\,\mathrm{GeV}^2$. Along the horizontal axis runs
$^{10}\log(\textrm{\# events})$. Two curves of the same type give the average
{\em plus} the standard deviation and the average {\em minus} the standard
deviation.}
\end{center}
\begin{center}
\begin{tabular}{|c|c|c||c|c|c|c|c|}
\hline
  \multirow{2}*{integrand}
 &\multirow{2}*{graph}
 &\multirow{2}*{algorithm}
     & $I      $ & $\stdev$ & $\Nge  $ & $\stdev^2\Ttot/\Tint$ & $\stdev^2\Nac$ \\
 & & & $10^{-8}$ & $\%    $ & $10^{3}$ & $10^{3}             $ & $10^{3}      $ \\\hline\hline
  \multirow{4}*{$\Asym$} 
 &\multirow{2}*{1}  &Metropolis& $1.594$ & $1.04 $ & $ 150$ & $0.0134$ & $0.0163 $ \\\cline{3-8}
 &                  &     \haag& $1.617$ & $0.713$ & $ 100$ & $0.0122$ & $0.00139$ \\\cline{2-8}
 &\multirow{2}*{2}  &    \parni& $1.528$ & $2.58 $ & $1000$ & $0.664 $ & $0.468  $ \\\cline{3-8}
 &                  &    \rambo& $1.647$ & $10.0 $ & $1000$ & $8.03  $ & $7.44   $ \\\hline\hline
  \multirow{4}*{$(\Asym)^{0.9}$} 
 &\multirow{2}*{3}  &             \sarge& $885.6$ & $0.866$ & $1000$ & $0.00424$ & $0.00264$ \\\cline{3-8}
 &                  &             \rambo& $823.9$ & $3.14 $ & $1000$ & $0.765  $ & $0.732  $ \\\cline{2-8}
 &\multirow{2}*{4}  &\parni, initialized& $889.4$ & $1.32 $ & $1000$ & $0.167  $ & $0.118  $ \\\cline{3-8}
 &                  &             \parni& $886.8$ & $1.69 $ & $1000$ & $0.273  $ & $0.198  $ \\\hline
\end{tabular}
\end{center}
The final results corresponding with \Figure{fig6}. $I$ is the integral,
$\stdev$ the standard deviation, $\nda$ the number of generated events, 
$\Ttot$ the total computation time, $\Tint$ the time it takes to perform one
evaluation of the integrand and $\Nac$ the number of accepted (non-zero weight)
events.  
\end{figure}

We apply all this to the problem of
calculating (\ref{antennaint}), with the function $\Ant$ replaced by a sum over
permutations of its arguments, so that the integrand looks more like a
$QCD$-amplitude: so we want to calculate
\begin{equation}
   \meas{\Asym} \df 
   \int d\Phi_n(Q;m_1^2,\ldots,m_n^2;p_1,\ldots,p_n)
        \,\Asym(p_0,p_1,\ldots,p_n,p_{n+1})
\;\;,
\label{antennasym}\end{equation}
with
\begin{displaymath}
   \Asym(p_0,p_1,\ldots,p_n,p_{n+1})
   = \sum_{\pi\in \mathit{Sym}(n+1)}
            \Ant(p_0,p_{\pi(1)},\ldots,p_{\pi(n)},p_{\pi(n+1)})
\;\;,
\end{displaymath}
and with $d\Phi_n$, $\Ant$ as in (\ref{phasespace}), (\ref{antenna}).
Below we put the integration process for the calculation of
$\meas{\Asym^{1-\alpha}}_{\Asym^\alpha}$ and the initialization process of 
\parni\ in a flow chart. The variables are
the set $p$ of $n$ momenta and $\dap\in[0,1]^{3n-4}$. 
\begin{align}
  \fbox{Metropolis with $\Asym^{\alpha}$}
  &\;\overset{\displaystyle p}{\longrightarrow}\;
  \fbox{\hicom\,$^{-1}$}\;\overset{\displaystyle\dap}{\longrightarrow}\;
  \fbox{\parni} 
  \nl
  \downarrow p& 
  \nl
  \fbox{integrand $\Asym^{1-\alpha}$}&
  \;\longrightarrow\;
  \fbox{$\textrm{sum}\lar\textrm{sum}+\Asym^{1-\alpha}(p)$}
\notag\end{align}
In the chart for the integration process of $\meas{\Asym^{\alpha}}$ and 
further optimization of \parni, also the
weight factors coming from \parni\ and \hicom\ are included:
\begin{align}
 &\fbox{\parni}\;\overset{\displaystyle\dap}{\longrightarrow}\;
  \fbox{\hicom}\;\overset{\displaystyle p}{\longrightarrow}\;
  \fbox{integrand $\Asym^{\alpha}$}
 \nl
 \nearrow
 &\hspace{20pt}w_{{\tt P}}\searrow
  \hspace{40pt}\downarrow w_{{\tt H}}
  \hspace{50pt}\swarrow\Asym^{\alpha}(p)
 \nl
  w_{{\tt P}}w_{{\tt H}}\Asym^{\alpha}(p)\;&\longleftarrow\;
  \fbox{$\textrm{sum}\lar\textrm{sum}+w_{{\tt P}}w_{{\tt H}}\Asym^{\alpha}(p)$}
\notag\end{align}
The results are presented in 
\Figure{fig5} 
for $n=4$, 
and in
\Figure{fig6} 
for $n=6$. 
The parameter $\alpha$ was put to $0.9$ for all cases.

We start the discussion with the results for $n=4$.
In order to obtain the curves with the title `Metropolis' in graph 1, the
calculation of $\meas{\Asym^{\alpha}}$ was done first with \sarge, so that
$\meas{\Asym}=\meas{\meas{\Asym^{\alpha}}\Asym^{1-\alpha}}_{\Asym^\alpha}$
could be calculated directly with the metropolis algorithm. In a realistic
calculation this is not necessary: this has only been done in order to arrive
at curves at the same scale as the curves from \haag\ and graph 2. The
Metropolis-curves converge rather quickly, as expected since the weights do not
fluctuate much because of the suppression from the exponentiation with
$1-\alpha$.

Graph 2 depicts the process of convergence in the direct calculation of 
$\meas{\Asym}$ with \parni\ and \rambo. 
The latter under estimates the integral for many events and hardly converges. 
\parni\ starts converging quickly after $10^5$ events, obviously after the 
optimization took place. 

The fact that the estimated errors with \rambo\ in graph 2 cannot be trusted
becomes clear from the curves in graph 3 for the calculation of
$\meas{\Asym^{\alpha}}$, which is supposed to be (slightly) easier for \rambo,
while the curves look worse. Graph 4 depicts the curves from \parni\ for the
calculation of $\meas{\Asym^{\alpha}}$, both with and without initialization
using the data from the metropolis calculation of
$\meas{\Asym^{1-\alpha}}_{\Asym^\alpha}$.  And clearly the curves with
initialization converge faster.

For $n=6$, the graphs look more-or-less the same: the curves from \rambo\ 
hardly converge or constitute an under estimation, while the curves from 
\parni\ {\em do} converge. Furthermore, the curves with initialization 
reach a given estimated error sooner than without initialization.

The tables coming with the figures speak for themselves, except maybe of the
last two columns. These give a measure of the computation time, ``normalized''
with respect to the reached error. The first of the last two columns gives this
number for the actual calculation, in units of the computation time of the
integrand. The last column gives this number extrapolated to the case that the
evaluation of the integrand would be much more expensive than the generation of
events. For \rambo, which can be considered to be a very cheap algorithm, these
numbers are close to each other%
\footnote{The odd situation with Metropolis for $n=6$, where the first number
is smaller than the second, is a result of the fact that a data point is 
re-used if a new one is not accepted, and does not add to the computation time.
}.
The conclusion that can be drawn from the tables is that \parni\ performs 
much better than \rambo, and that \parni\ performs better with initialization
than without. The reason why \haag\ and \sarge\ perform much better than the
rest is that these algorithms have been specially designed to integrate 
exactly $\Asym$.


\newpage
\section{Summary}
A histogram has been shown to be a special case of the application of the {\em
multi-channel\/} method to derive a probability distribution from a sample, or
a stream, of data. The channels are the normalized bins of the histogram, and
the channel weights are the volumes of the bins. An algorithm has been
presented how to optimize the weights in the case of channels that consist of
arbitrary probability densities, leading to a {\em unitary probability decomposition} (\upd).

Furthermore, the algorithm \parni\ has been presented, for which not only the
channel weights, but also the channels themselves are adaptive. It can be used
both for the creation of \upds\ and for automatic importance sampling in Monte
Carlo integration. It handles data in an $\dimO$-dimensional hypercube
$[0,1]^\dimO$ for arbitrary $\dimO$. 
It has no \apriori\ problems with non-factorizable peak structures in the 
integrand, and its complexity grows linearly with $\dimO$.

Finally, it has been shown how the property of \parni\ to create a \upd\ can be
used to optimize the Monte Carlo integration procedure, and this has been 
applied to a problem in phase space integration.

\subsubsection*{Acknowledgment}
This research has been supported by a Marie Curie Fellowship of the European
Community program ``Improving Human Research Potential and the Socio-economic
Knowledge base'' under contract number HPMD-CT-2001-00105.

\newcommand{\EPJ}[3]{Eur.\ Phys.\ J.\ {\bf #1} (#2) #3}
\newcommand{\CPC}[3]{Comp.\ Phys.\ Comm.\ {\bf #1} (#2) #3}
\newcommand{\JCP}[3]{J.\ Comp.\ Phys.\ {\bf #1} (#2) #3}
\newcommand{\PL}[3]{Phys.\ Lett.\ {\bf #1} (#2) #3}
 
%

\section{Appendices}
\renewcommand{\thesubsection}{\Alph{subsection}}
\renewcommand{\thesubsubsection}{\thesubsection\arabic{subsubsection}}
\renewcommand{\theparagraph}{\thesubsubsection.\alph{paragraph}}
\newcommand{\countpar}{\refstepcounter{paragraph}}
\newcommand{\parcount}{\theparagraph\hspace{10pt}}

\subsection{Multi-channeling for importance sampling\label{VarOpt}}
In \cite{KleissPittau}, the multi-channel method was constructed such that the
variance of the Monte Carlo estimator of the integral of a function is
minimal.
We briefly repeat the line of argument.
A sample of data points
$\Dap=(\dap_1,\dap_2,\ldots,\dap_{\nda})$
is generated distributed following the density
$\Gdns{\we}$,
and the integral of integrand
$\Pdns$
over
$\dspace$
is estimated by
\begin{displaymath}
   \Omeas{\grac{\Pdns}{\Gdns{\we}}} \df
   \frac{1}{\nda}\sum_{i=1}^{\nda}\frac{\Pdns(\dap_i)}{\Gdns{\we}(\dap_i)}
   \overset{\nda\rightarrow\infty}{\longrightarrow} 
   \int_{\dspace}\Pdns(\dap)\,d\dap
\;\;.   
\end{displaymath}
The variance of the estimator is given by
\begin{equation}
  \frac{1}{\nda}
  \left( \int_{\dspace}\frac{\Pdns(\dap)^2}{\Gdns{\we}(\dap)}\,d\dap
        -\Big(\int_{\dspace}\Pdns(\dap)\,d\dap\Big)^2\right)
\;\;,
\label{variance1}\end{equation}
and extremization leads to the solution that the quantities
\begin{displaymath}
    W_i(\Pdns,\Gdns{\we})
    \df
    \int_{\dspace}\frac{\gdns_i(\dap)\Pdns(\dap)^2}{\Gdns{\we}(\dap)^2}\,d\dap
\end{displaymath}
have to be equal for all
$i=1,\ldots,\nwe$.
If
$\Gdns{\we}$
is a histogram, then the solution is immediately found using \Algorithm{alg1}
with the first step replaced by 
\begin{algorithm}[importance sampling by variance optimization]
\begin{enumerate}
\item $y_i \lar \we_i\sqrt{W_i(\Pdns,\Gdns{\we})}$
      \hspace{10pt}for all $i=1,\ldots,\nwe$
\end{enumerate}
\label{alg2}\end{algorithm}
which is also applied in the general case.
Of course the 
$W_i(\Pdns,\Gdns{\we})$
cannot be calculated exactly, but can be estimated by
$\Omeas{\grac{\gdns_i\Pdns^2}{\Gdns{\we}^3}}$,
since 
$\Dap$ 
is distributed following
$\Gdns{\we}$.

\subsection{HICOM\label{hicom}}
\newcommand{\Ss}{s}
\newcommand{\sa}{\sigma}
In the HIerarchical Construction Of Momenta, one uses the fact that the
phase space can be decomposed as
\begin{align}
  d\Phi_n(P;\sa_1,\ldots,\sa_n;p_1\ldots,p_n) 
  &= 
  d\Ss_{n-1}\,d\Phi_2(P;\sa_{n},\Ss_{n-1};p_n,Q_{n-1}) \nl
  &\times
  d\Ss_{n-2}\,d\Phi_2(Q_{n-1};\sa_{n-1},\Ss_{n-2};p_{n-1},Q_{n-2})\nl
  &\hspace{20pt}\vdots\label{decomposition}\\
  &\times 
  d\Ss_2\,d\Phi_2(Q_3;\sa_3,\Ss_2;p_3,Q_2)\;d\Phi_2(Q_2;\sa_2,\sa_1;p_2,p_1)
\;\;,
\notag\end{align}
where
\begin{displaymath}
  d\Phi_2(Q;s_1,s_2;q_1,q_2)
   \df 
    d^4q_1\,\dirac(q_1^2-s_1)\theta(q_1^0)
  \,d^4q_2\,\dirac(q_2^2-s_2)\theta(q_2^0)\,\delta^4(Q-q_1-q_2)
\end{displaymath}
is the standard two-body phase space. This decomposition tells us that if the
variables involved are generated in this order and with these dependencies, then
the final momenta are distributed on the desired bounded $(3n-4)$-dimensional 
subspace of $\Real^{4n}$. It does not tell us how the momenta are distributed, 
and there is the freedom how to generate the variables in each of the two-body
phase spaces, and the variables $\Ss_i$. Examples of particular choices to 
obtain momenta that are distributed following the antenna pole structure can 
be found in \cite{vanHamerenPapadopoulos}. 

A well-known example how to generate each of the two-body phase spaces is by
generating an angle $\vhi$ uniformly in $[0,2\pi]$, and a variable $z$ 
distributed uniformly in $[-1,1]$, performing the construction
\begin{align}
   |\vec{q}_1| &\lar \sqrt{\lambda(Q^2,\Ss_1,\Ss_2)/4/Q^2}
   \nl
   q_1^0 &\lar \sqrt{s_1+|\vec{q}_1|^2} 
   \nl
   \vec{q}_1 
   &\lar |\vec{q}_1|
        \big(\,\sqrt{1-z^2}\,\cos\vhi\,,\,\sqrt{1-z^2}\,\sin\vhi\,,\,z\,\big)
\;\;,
\notag\end{align}
boosting $q_1$ to the center-of-mass frame of $Q$, and putting $q_2=Q-q_1$.
The symbol $\lambda$ stands for the standard K\"allen function. 
This construction gives a Jacobian factor
$2Q^2/\pi/\sqrt{\lambda(Q^2,s_1,s_2)}$ in the density. If we look at the
decomposition (\ref{decomposition}) in the case that all squared masses 
$\sa_i$ are zero, we see that the Jacobian factors are equal to 
$2Q_{i}^2/\pi/(Q_{i}^2-\Ss_{i-1})$. Since $Q_i^2=\Ss_i$ in the end, we see that
the non-constant parts of the Jacobian factors cancel if the variables $\Ss_i$
are generated following the density
\begin{displaymath}
                    i(i-1)\frac{(\Ss_{i+1}-\Ss_{i})(\Ss_{i})^{i-2}}
                            {(\Ss_{i+1})^{i}} 
\;\;,
\end{displaymath}
leading to momenta that are uniformly distributed over phase space with 
constant density
\begin{displaymath}
   \frac{(n-1)(n-2)\,2P^2}{(P^2)^{n-1}\,\pi}\cdot\frac{(n-2)(n-3)2}{\pi}
   \cdots\frac{(2)(1)2}{\pi}\cdot\frac{2}{\pi}
   = \left(\frac{2}{\pi}\right)^{n-1}\frac{\Gamma(n)\Gamma(n-1)}{(P^2)^{n-2}}
\;\;.
\end{displaymath}
The density for the variable $\Ss_{i}$ is obtained by generating $\rho\in[0,1]$
following the {\em beta\/}-density
$\beta_{i-1,2}(\rho)=i(i-1)(1-\rho)\rho^{i-2}$,
and putting $\Ss_i\lar\Ss_{i+1}\rho$. 

Cheng's BA algorithm \cite{Devroye} is 
very efficient in generating random numbers following 
{\em beta\/}-distributions. However, it uses the method of rejection and needs,
on the average, more than one uniformly distributed random number for 
returning one {\em beta\/}-variable. 
What we would like is a direct construction with one random number as input, and
the {\em beta\/}-variable as output, so that we can let \parni\ deliver the 
input. Fortunately, Cheng's BA algorithm is so efficient that we can skip the
rejection part, and just use its construction of a variable that is 
almost a {\em beta\/}-variable. If necessary, \parni\ will compensate for that. 
For a general {\em beta\/}-density 
\begin{displaymath}
   \beta_{a,b}(\rho) \propto (1-\rho)^{b-1}\rho^{a-1}
\;\;,
\end{displaymath}
The density for the almost-{\em beta\/}-variable is given by 
\begin{displaymath}
   \frac{d}{d\rho}\,\frac{\rho^u}{\rho^u + (a/b)^u(1-\rho)^u}
\;\;,
\end{displaymath}
with
\begin{displaymath}
   u = \begin{cases}
           \min(a,b) &\textrm{if $\min(a,b)\leq1$}\\
           \sqrt{(2ab-a-b)/(a+b-2)} &\textrm{if $\min(a,b)>1$}
       \end{cases}
\;\;.
\end{displaymath}

\end{document}